\documentclass[preprint,12pt]{elsarticle}

\usepackage{amssymb}
\usepackage{amsmath}

\usepackage{placeins}

\usepackage{times}

\usepackage{amsmath}
\usepackage{amsfonts}
\usepackage{amssymb}
\usepackage{amsthm}
\usepackage{dsfont}
\usepackage{mathtools}
\usepackage{bbm}

\usepackage{siunitx}

\usepackage{marvosym}

\usepackage{graphicx}
\usepackage{enumerate}
\usepackage{hyperref}
\usepackage[ruled,longend]{algorithm2e}
\usepackage{fancyref}
\usepackage{etoolbox}
\usepackage{appendix}
\usepackage{soul}
\usepackage[dvipsnames]{xcolor}
\usepackage[noabbrev,capitalise]{cleveref}
\usepackage{colortbl}
\hypersetup{colorlinks,linkcolor=blue,citecolor=red,urlcolor=blue}

\creflabelformat{equation}{#2#1#3}

\renewcommand{\vec}[1]{\boldsymbol{#1}}
\newcommand{\mat}[1]{\mathbf{#1}}
\renewcommand{\eqref}[1]{eq.~(\ref{#1})}
\newcommand{\figref}[1]{Figure~\ref{#1}}
\newcommand{\subfigref}[2]{Figure~\hyperref[#1]{\ref*{#1}#2}}
\newcommand{\tabref}[1]{Table~\ref{#1}}

\newcommand{\edit}[1]{{\color{OliveGreen}{#1}}}
\renewcommand{\edit}[1]{#1}

\journal{Computer Methods in Applied Mechanics and Engineering, }

\begin{document}

\begin{frontmatter}



\title{Tensor Network Lattice Boltzmann Method for Data-Compressed Fluid Simulations}

\author[aff1,equal,corresponding]{Lukas Gross}
\author[aff1]{Elie Mounzer}
\author[aff2,equal]{David M. Wawrzyniak}
\author[aff3]{Josef M. Winter}
\author[aff2,aff3]{Nikolaus A. Adams}


\affiliation[aff1]{organization={German Research Center for Artificial Intelligence, Robotics Innovation Center},
            addressline={Robert-Hooke-Str. 1},
            city={Bremen},
            postcode={28359},
            state={},
            country={Germany}}

\affiliation[aff2]{organization={Technical University of Munich, Munich Institute of Integrated Materials, Energy and Process Engineering},
            addressline={Lichtenbergstr. 4a},
            city={Garching},
            postcode={85748},
            state={},
            country={Germany}}

\affiliation[aff3]{organization={Technical University of Munich, School of Engineering and Design, Chair of Aerodynamics and Fluid Mechanics},
            addressline={Boltzmannstr. 15},
            city={Garching},
            postcode={85748},
            state={},
            country={Germany}}

\fntext[equal]{These authors contributed equally to this work.}
\fntext[corresponding]{Corresponding author, email: lukas.gross@dfki.de}
\begin{abstract}
Resolving unsteady transport phenomena in geometrically complex domains is traditionally constrained by polynomial scaling of computational cost with spatial resolution. While methods based on tensor-network data representations or matrix-product states (MPS) data encodings have emerged as a technique to systematically reduce degrees of freedom, existing formulations do not extend to complex geometries and complex flow physics. Both capabilities are offered by lattice Boltzmann methods, for which we develop a generalized MPS formulation. This development marks a paradigm shift from classical methods that rely on explicit grid refinement for data reduction. Instead, our approach exploits non-local correlations in the MPS representation to systemically compress the global fluid state directly without modifying the underlying grid. We benchmark the proposed solver against classical LBM using three-dimensional flows through structured media and vascular geometries. The results confirm that the MPS formulation reproduces the reference solution with high fidelity while achieving compression ratios exceeding two orders of magnitude, positioning tensor networks or MPS encodings as a scalable paradigm for continuum mechanics on high-performance GPU hardware.

\end{abstract}



\begin{keyword}
lattice Boltzmann method, matrix product states, CFD

\end{keyword}

\end{frontmatter}


\section{Introduction}
In computational fluid dynamics (CFD), the numerical solution to the Navier-Stokes equations (NSE) is key to dealing with engineering problems and for investigating physical phenomena. Complex flows involve a broad range of scales that need to be resolved directly or approximated. With direct numerical simulation (DNS) the computational mesh resolution is sufficiently high to resolve the smallest possible flow scales, independent of internal scale relations. DNS implies extreme computational effort, which can be reduced by multi-resolution strategies, but nevertheless requires the use of high-performance computing systems. Some of the classical strategies for DNS are gradient-based adaptive mesh refinement~\cite{berger_local_1989} and wavelet-based multiresolution algorithms~\cite{harten_multiresolution_1995,han_adaptive_2014,kaiser_multiresolution_2021,hoppe_parallel_2022}.
Mesh-independent data compression methods have found new recent interest with the quantum-inspired introduction of matrix product states (MPS)~\cite{schollwock_density-matrix_2011, oseledets_tensor-train_2011, orus_practical_2014}, and the density matrix renormalization group (DMRG)~\cite{white_density_1992} for constituted state encodings with inherent error control.
DMRG and its MPS variants belong to a class of algorithms, originally developed for simulations of many-body quantum systems with small entanglement~\cite{vidal_entanglement_2007, Verstraete_TN_2008}.
Gourianov et al.~\cite{gourianov_quantum-inspired_2022} have introduced an algorithm based on MPS decomposition and a DMRG-like approach to solve a variational formulation of the incompressible NSE, drawing an analogy between spatial correlations of quantum states under local Hamiltonians~\cite{eisert_area-laws_2010} and interscale correlations of turbulent flows. 
Several extensions have since been published. 
Kiffner and Jaksch~\cite{kiffner_tensor_2023} introduced resting and moving wall boundary conditions and proposed a more explicit algorithm using a DMRG-like routine~\cite{oseledets_solution_2012} to solve the Poisson equation. 
Peddinti et al.~\cite{peddinti_quantum-inspired_2024} proposed methods to treat immersed solid boundaries and for efficient evaluation of solutions while using a similar DMRG-like solver for the Poisson equation. 
An extension to curvilinear grids was proposed by van Hülst et al.~\cite{hulst_quantum-inspired_2025}, and an implementation for GPUs was proposed by Hölscher et al.~\cite{holscher_quantum-inspired_2025}. Moreover, Ghahremani and Babaee~\cite{ghahremani_cross_2024} utilized cross interpolation for high-dimensional dynamical systems. 
Thai Tran et al.~\cite{thai_tran_tensor_2026} developed an MPS-based\footnote{Thai Tran et al. use the term tensor train, which is another name for the same form of tensor decomposition.} isogeometric solver for large-scale 3D Poisson problems.
MPS-based approaches have so far been developed primarily for data compression of turbulent flows in relatively simple geometries, leaving their applicability to complex domains and non-turbulent flows largely unexplored.

Known for its capabilities of representing geometrically and physically complex fields, the lattice Boltzmann method (LBM) is a widely adopted approach for solving the NSE.
In this framework, discrete velocity sets are employed to represent single-particle distribution functions.
A low Mach-number approximation is achieved through a multiscale expansion, where the collision term can be formulated as a linear relaxation towards an equilibrium using the Bhatnagar–Gross–Krook (BGK) model~\cite{bhatnagar_model_1954}.
This term accounts for local non-equilibrium effects, thereby introducing nonlinearity through an approximated local equilibrium distribution.
Compared to directly solving the NSE, LBM exhibits algorithmic simplicity through operator splitting between linear streaming and nonlinear collision, at the cost of higher dimensionality than typical finite-difference or finite-volume schemes.
In two dimensions, LBM typically evolves nine distinct distribution functions, while in three dimensions it requires at least fifteen, making it a memory-intensive approach.
The advection of distribution functions is inherently linear and exact, and implementing complex boundary conditions within the LBM framework is comparably easy.
Moreover, the method often employs a uniform computational grid.
Although LBM, as a memory-intensive method, offers high potential for data compression, few compression techniques have been developed for it~\cite{dupuis_theory_2003,eitel-amor_lattice-boltzmann_2013,liu_efficient_2022}, and none of them are independent of the discretization mesh.
A major challenge arises from the coupling between spatial and temporal discretization on uniform grids, which prevents straightforward adaptation of local grid refinement and requires special treatment~\cite{bellotti_multidimensional_2022}.
These characteristics make LBM an excellent candidate for MPS data encoding.
Using the MPS formulation, we introduce a computational framework where data compression is achieved through low-rank approximation of non-local correlations, without manipulation of the underlying grid.
The substantial memory demand associated with storing distribution functions can thus be effectively mitigated through MPS compression while preserving the explicit weakly compressible formulation of LBM to circumvent the need to solve a Poisson equation.
Furthermore, the uniform grid structure intrinsic to LBM is naturally compatible with the MPS representation.

\begin{figure}[htb]
    \centering
    \includegraphics[width=\linewidth]{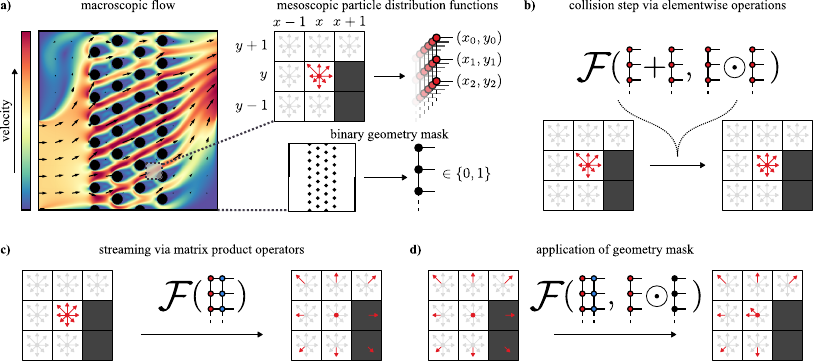}
    \caption{\textbf{Overview of the tensor-network lattice Boltzmann method.} \textbf{a)} The macroscopic flow is represented by mesoscopic particle distribution functions. Each \edit{particle distribution function} is individually decomposed into a tensor network or matrix product state (MPS). The geometry is defined by a binary object mask, which is also decomposed into an MPS of the same shape as the \edit{particle distribution functions}. \textbf{b)} The collision step models local particle interactions through element-wise operations. These are rewritten in order to be efficiently executed in the MPS-manifold via element-wise summation and multiplication. \textbf{c)} The streaming step propagates the \edit{particle distribution functions} in their respective directions. In the MPS-manifold this step is realized through exact low-rank matrix product operators (MPO). \textbf{d)} Boundary conditions at complex geometries are enforced by the bounce-back scheme. MPS-LBM encodes boundaries in the MPS manifold by masking the \edit{particle distribution functions} and applying additional streaming to masked regions.}
    \label{fig:overview}
\end{figure}
We propose a novel MPS-based method for solving LBM (MPS-LBM). We overcome limitations of previous quantum-inspired CFD approaches by exploiting the flexible formulation of LBM to realize MPS-compressed simulations in complex domains.
Decomposing distribution functions as MPS enables high compression rates while maintaining excellent accuracy.
Efficient low-rank matrix product operators (MPO) are employed for the streaming step.
See \figref{fig:overview} for a graphical overview.
Additionally, we implement both temporally varying inflow boundary conditions as well as pressure-based outflow conditions.
Despite these extensions, MPS-LBM maintains logarithmic scaling in the spatial resolution and quartic scaling in bond dimension, which is characteristic of quantum-inspired CFD methods.
The proposed algorithm is validated on the example of the Taylor-Green vortex in three dimensions.
Furthermore, a three-dimensional simulation of blood flow in a realistic aneurysm geometry illustrates a typical application of LBM in engineering applications. Finally, an industrially relevant case of a geometrically complex flow is presented. Additional test cases are provided in \ref{sup:2dcases}, demonstrating the full capabilities of the proposed algorithm.
We show that MPS-LBM for the complex geometry case exhibits exceptionally high compression while maintaining high accuracy, highlighting the utility of MPS decompositions in complex flows around such geometries.

The remainder of this paper is organized as follows. 
The fundamentals of the MPS formalism are introduced in \cref{sec:mps}, focusing on specific aspects of MPS-LBM. This section explicitly derives key algorithmic operations: the scale ordering transformation, addition and multiplication, and the closed-form Matrix Product Operator (MPO) for the streaming step.
\cref{sec:mpslbm} establishes the theoretical framework of the Lattice Boltzmann Method, detailing the governing equations as well as their integration with the MPS formalism. It also details the implementation of boundary conditions within the MPS framework. 
\cref{sec:setup} defines the numerical benchmarks, specifically the three-dimensional Taylor-Green vortex, blood flow through an aneurysm, and a pin-fin array, and the error metrics used for validation. 
\cref{sec:results}, we present a quantitative comparison between the proposed MPS-LBM and a reference classical LBM implementation. Finally, \cref{sec:discussion} discusses the implications of these results for scalable fluid simulation in complex geometries.

\section{The Matrix Product State Formalism}\label{sec:mps}

\subsection{Compression with Matrix Product States}
\begin{figure}
    \centering
    \includegraphics[width=0.99\linewidth]{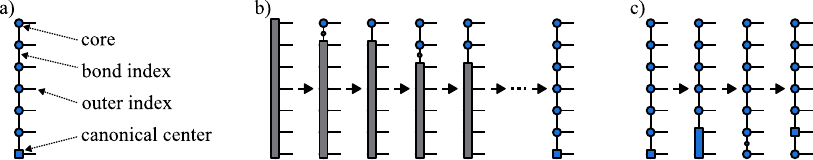}
    \caption{\textbf{Diagrammatic explanation of Matrix Product States.} \textbf{a)} Tensor network diagram of an MPS with annotations. \textbf{b)} Decomposition of a larger tensor into an MPS as a tensor network diagram. \textbf{c)} Shift of the canonical center of an MPS as a tensor network diagram}
    \label{fig:mpsintro}
\end{figure}

An alternative to grid-based data encodings is given by Matrix Product States (MPS)~\cite{schollwock_density-matrix_2011, oseledets_tensor-train_2011, orus_practical_2014}. An MPS is a specific type of tensor decomposition that allows effective compression and efficient manipulation of large tensors. Specifically, let $\mat{T}^{\omega_1,\dots,\omega_n}$ be an order-$n$ tensor, then there exists a set of tensors $\mat{A}_i$ such that
\begin{equation}
    \mat{T}^{\omega_1,\dots,\omega_n} = \sum_{\{\vec{\alpha}\}} \left(\mat{A}_1\right)_{\alpha_1}^{\omega_1} \left(\mat{A}_2\right)_{\alpha_1, \alpha_2}^{\omega_2} \dots \left(\mat{A}_n\right)_{\alpha_{n-1}}^{\omega_n}.
    \label{eq:MPS}
\end{equation}
The tensor $\left(\mat{A}_i\right)$ is called \emph{core} at \emph{site} $i$. The maximum dimension $\chi$ of any \emph{bond} index $\alpha_l$ is called \emph{bond dimension} of the MPS, i.e. $\chi=\max_l|\alpha_l|$. It is common to depict these kinds of tensor decompositions in a diagrammatic language, where basic geometric shapes represent a tensor and a link between shapes represents a common index that is summed over (see \subfigref{fig:mpsintro}{a}). An MPS can be constructed iteratively from a larger tensor via a sequence of singular value decompositions (SVD). It starts by reshaping $\mat{T}$ to a matrix, then applying SVD, contracting the singular values $\mat{S}$ into the right resulting matrix $\mat{B}$, and repeating the same on $\mat{SB}$. In the following, we will denote independent indices as $\alpha,\beta,...$ and multiple indices combined into a single one as $(\alpha\beta...)$.
\begin{equation}
    \begin{aligned} 
        \mat{T}^{\omega_1,(\omega_2\dots\omega_n)} \xrightarrow[]{\text{SVD}} &\sum_{\alpha_1} (\mat{A}_1)^{\omega_1}_{\alpha_1} (\mat{S}_1)_{\alpha_1} (\mat{B}_1)^{(\omega_2\dots\omega_n)}_{\alpha_1} \\
        \xrightarrow[]{\text{contract and reshape}} &\sum_{\alpha_1} (\mat{A}_1)^{\omega_1}_{\alpha_1} (\mat{S}_1\mat{B}_1)^{(\alpha_1\omega_2),(\omega_3\dots\omega_n)} \\
        \xrightarrow[]{\text{SVD}} &\sum_{\alpha_1, \alpha_2} (\mat{A}_1)^{\omega_1}_{\alpha_1} (\mat{A}_2)^{\omega_2}_{\alpha_1,\alpha_2} (\mat{S}_2)_{\alpha_2} (\mat{B}_2)^{(\omega_3\dots\omega_n)}_{\alpha_2} \\
        \xrightarrow[]{\text{contract and reshape}} &\sum_{\alpha_1, \alpha_2} (\mat{A}_1)^{\omega_1}_{\alpha_1} (\mat{A}_2)^{\omega_2}_{\alpha_1,\alpha_2} (\mat{S}_2\mat{B}_2)^{(\alpha_2\omega_3),(\omega_4\dots\omega_n)} \\
        \dots \\
        \xrightarrow{} &\sum_{\{\vec{\alpha}\}} \left(\mat{A}_1\right)_{\alpha_1}^{\omega_1} \left(\mat{A}_2\right)_{\alpha_1, \alpha_2}^{\omega_2} \dots \left(\mat{C}_n\right)_{\alpha_{n-1}}^{\omega_n},
    \end{aligned}
    \label{eq:svd-comp}
\end{equation}
where $\mat{C}_n = \mat{S}_n\mat{B}_n$. See also \subfigref{fig:mpsintro}{b}. Truncating singular values at each step allows to reduce the number of elements of the $\mat{A}_i$, thus compressing the data encoded in $\mat{T}$.
SVD ensures that upon truncation to a certain rank the truncation error is minimal in the $\text{l}^2$-norm at each step and scales with the largest omitted singular value. Since all $\mat{A}_i$ are left singular matrices they are all left-orthogonal i.e. 
\begin{equation}
    \sum_{\alpha_, \omega_i} \left(\mat{A}_i\right)_{\alpha, \beta}^{\omega_i} \left(\mat{A}_i\right)_{\alpha, \gamma}^{\omega_i} = \mathbb{I}_{\beta,\gamma}.
    \label{eq:lorg}
\end{equation}
Because of this the squared norm of $\mat{T}$ reduces to
\begin{equation}
    \sum_{\{\omega\}}\mat{T}^{\omega_i, \dots, \omega_n}\mat{T}^{\omega_i, \dots, \omega_n} = \sum_{\alpha_{n-1}, \omega_n} \left(\mat{C}_{n}\right)_{\alpha_{n-1}}^{\omega_n}\left(\mat{C}_{n}\right)_{\alpha_{n-1}}^{\omega_n},
\end{equation}
where the core at site $n$ is called \emph{canonical center}. This canonical center can be shifted to the left using SVD:
\begin{equation}
    \begin{aligned}
        & \dots \left(\mat{A}_{n-1}\right)_{\alpha_{n-2},\alpha_{n-1}}^{\omega_{n-1}} \left(\mat{C}_{n}\right)_{\alpha_{n-1}}^{\omega_n} \\
        \xrightarrow[]{\text{contract and reshape}} & \dots \left(\mat{A}_{n-1}\mat{C}_n\right)^{(\alpha_{n-2}\omega_{n-1}),(\omega_n)} \\
        \xrightarrow[]{\text{SVD}} & \dots \left(\mat{A'}_{n-1}\right)_{\alpha_{n-2},\alpha_{n-1}}^{\omega_{n-1}} \left(\mat{S}\right)_{\alpha_n-1} \left(\mat{B}_{n}\right)_{\alpha_{n-1}}^{\omega_n} \\
        \xrightarrow[]{\text{contract and reshape}} & \dots \left(\mat{C}_{n-1}\right)_{\alpha_{n-2},\alpha_{n-1}}^{\omega_{n-1}} \left(\mat{B}_{n}\right)_{\alpha_{n-1}}^{\omega_n},
    \end{aligned}
    \label{eq:svd-move}
\end{equation}
where $\mat{C}_{n-1} = \mat{A'}_{n-1}\mat{S}$. See also \subfigref{fig:mpsintro}{c}. Analogously to \cref{eq:lorg}, $\mat{B}_n$ is now right-orthogonal and thus
\begin{equation}
    \sum_{\{\omega\}}\mat{T}^{\omega_i, \dots, \omega_n} = \sum_{\alpha_{n-2},\alpha_{n-1}, \omega_{n-1}} \left(\mat{C}_{n-1}\right)_{\alpha_{n-2},\alpha_{n-1}}^{\omega_{n-1}}.
    \label{eq:ccenter}
\end{equation}
Similar to \cref{eq:svd-comp}, this operation can be performed iteratively, and the singular values can again be truncated to achieve additional compression. Generally, compression should only be performed on the canonical center, as from \cref{eq:ccenter} follows that the canonical center encodes all information with respect to the $\text{l}^2$-norm of the entire MPS. A truncated SVD on a site other than the canonical center produces a locally valid compression but degrades the global accuracy of the compression. For more details on different construction approaches as well as further methods to compress an MPS, we refer to~\cite{schollwock_density-matrix_2011, oseledets_tensor-train_2011, orus_practical_2014}.

\subsection{Algebra on Matrix Products States}
\label{sec:mpsalgebra}
In addition to compression, the MPS formalism allows for manipulation of data within the compressed representation without the necessity to contract all cores. Many operations can be efficiently performed in a site-wise manner. In the following we consider two tensors $\mat{A}$ and $\mat{B}$ with respective MPS representations with cores $(\mat{A}_i)^{\omega_i}_{\alpha_{i-1}, \alpha_i}$ \edit{and} $(\mat{B}_i)^{\omega_i}_{\beta_{i-1}, \beta_i}$. Elementwise addition~\cite{schollwock_density-matrix_2011, oseledets_tensor-train_2011, orus_practical_2014} of $\mat{A}$ and $\mat{B}$ is achieved by combining the respective cores into block matrices as
\begin{equation}
    \left((\mat{A} + \mat{B})_i\right)^{\omega_i}_{\alpha_{i-1}, \alpha_i} = 
    \begin{pmatrix}
        \mat{A}_i^{\omega_i} & 0 \\
        0 & \mat{B}_i^{\omega_i}
    \end{pmatrix}_{\alpha'_{i-1}, \alpha'_i},
    \label{eq:add}
\end{equation}
where the new indices $\alpha'_i$ append the entries of $\beta_i$ to $\alpha_i$ and therefore $\left|\alpha'_i\right| = \left|\alpha_i\right| + \left|\beta_i\right|$. To perform elementwise multiplication, the respective pairs of cores are contracted with a Kronecker delta tensor $\delta^{i,j,k}$, which has entries $1$ if $i=j=k$ and $0$ otherwise.
\begin{equation}
    \left((\mat{A} \odot \mat{B})_i\right)^{\omega'_i}_{\alpha'_{i-1}, \alpha'_i} = \sum_{l,m} \delta^{\omega'_i,l,m} \left(\mat{A}_i\right)^{l}_{\alpha_{i-1}, \alpha_i} \left(\mat{B}_i\right)^{m}_{\beta_{i-1}, \beta_i},
    \label{eq:mul}
\end{equation}
where $\alpha'_i = (\alpha_i\beta_i)$ and thus the bond dimensions are multiplied, $\left|\alpha'_i\right| = \left|\alpha_i\right|\left|\beta_i\right|$. The equivalent of a linear operator mapping tensors to tensors is a Matrix Product Operator (MPO). Compared to an MPS, an MPO has two outer indices at each site but otherwise the same structure.
\begin{equation}
    \mat{O}^{\nu_1,\dots,\nu_n,\omega_1,\dots,\omega_n} = \sum_{\{\vec{\gamma}\}} \left(\mat{O}_1\right)_{\gamma_1}^{\nu_1,\omega_1} \left(\mat{O}_2\right)_{\gamma_1, \gamma_2}^{\nu_2,\omega_2} \dots \left(\mat{O}_n\right)_{\gamma_{n-1}}^{\nu_n,\omega_n}.
    \label{eq:MPO}
\end{equation}
An MPO is applied to an MPS site-wise by contracting one of the MPO sets of physical indices with the physical indices of the MPS, resulting again in an MPS.
\begin{equation}
    \left((\mat{O}\mat{A})_i\right)^{\nu_i}_{\gamma'_{i-1}, \gamma'_i} = \sum_{\omega_i} \left(\mat{O}_i\right)^{\nu_i, \omega_i}_{\gamma_{i-1}, \gamma_i} \left(\mat{A}_i\right)^{\omega_i}_{\alpha_{i-1}, \alpha_i},
    \label{eq:applympo}
\end{equation}
where the bond dimensions multiply aigain $\left|\gamma'_i\right| = \left|\gamma_i\right|\left|\alpha_i\right|$.

All of the above operations increase the bond dimensions. To maintain compression during manipulations, it is necessary to control bond dimensions $\chi$. The naive approach to iteratively apply \cref{eq:svd-move} with truncation scales as $\mathcal{O}(\chi^3)$, which is feasible for the mild increase upon a single addition operation or the application of MPO of low bond dimension. However, for the multiplication of two MPS of similar bond dimension $\chi$, this naive approach scales as $\mathcal{O}(\chi^6)$, which demands more efficient methods as discussed below.

\subsection{Compressed Elementwise Multiplication}\label{sec:mul}
\begin{figure}[htb]
    \centering
    \includegraphics[width=0.99\linewidth]{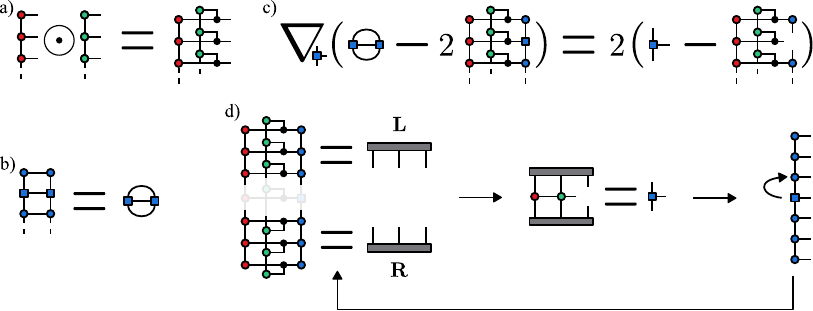}
    \caption{\textbf{Diagrammatic explanation of compressed element-wise multiplication.} The product MPS to be multiplied is depicted with red and green sites, the candidate MPS has blue sites, and small black sites are Kronecker delta tensors. \textbf{a)} Diagram for exact element-wise multiplication. Physical dimensions of each MPS site are connected via Kronecker delta tensors (black dots). \textbf{b)} Orthogonality conditions simplify the inner product of MPS to the inner product of canonical centers. \textbf{c)} Local gradient with respect to the candidate canonical center of the term to be minimized. \textbf{d)} Iteration loop: First the left $L$ and the right side $R$ of the tensor network are contracted. The sites of the product MPS corresponding to the candidate canonical center are contracted with $R$ and $L$ to form the new canonical center. The new center is inserted in the candidate, and the canonical center is shifted.}
    \label{fig:multiplication}
\end{figure}

We detail the algorithm employed for element-wise multiplication with simultaneous compression, the single-site version of the algorithm proposed in~\cite{verstraete_mult_2004}. We also briefly discuss possible alternatives. Consider the element-wise product of two MPS $\mat{B}$ and $\mat{C}$ and a candidate MPS $\mat{A}$, that has the desired bond dimension. $\mat{A}$ is constructed to approximate $\mat{B}\odot\mat{C}$ by minimizing the squared $\text{l}^2$-norm
\begin{equation}
    \begin{aligned}
        &\min_{\mat{A}} \left\lVert \mat{A} - (\mat{B} \odot \mat{C}) \right\rVert_2^2, \\
        \Leftrightarrow &\nabla_{\mat{A}}\left\lVert \mat{A} - (\mat{B} \odot \mat{C}) \right\rVert_2^2 = 0, \\
        \Leftrightarrow &\nabla_{\mat{A}} \left( \mat{A} \cdot \mat{A}\right) - \nabla_{\mat{A}} \left(2\mat{A} \cdot (\mat{B} \odot \mat{C}) \right) + \underbrace{\nabla_{\mat{A}} \left((\mat{B} \odot \mat{C}) \cdot (\mat{B} \odot \mat{C}) \right)}_{=0}= 0.
    \end{aligned}\label{eq:minimization}
\end{equation}
Following \cref{eq:mul}, we can write the element-wise product as the tensor network diagram shown in
\subfigref{fig:multiplication}{a}, where black dots denote Kronecker delta tensors. Now let $\mat{A}$ be in canonical form with the center at site $c$.
Then $\mat{A} \cdot \mat{A}$ reduces to the scalar product of $\mat{A}_c$ with itself as in \subfigref{fig:multiplication}{b}. Instead of the gradient with respect to the whole of $\mat{A}$ in \cref{eq:minimization}, we consider the local gradient at site $c$, and derive the respective tensor network as in \subfigref{fig:multiplication}{c}. The local minimum is found by setting $\mat{A}_c$ equal to the contraction of the right-most term in \subfigref{fig:multiplication}{c}. Minimization is achieved iteratively as in \subfigref{fig:multiplication}{d}: Contract the tensor network, replace the canonical center, shift the canonical center, and repeat. The contraction of the tensor network scales as $\mathcal{O}(n\chi^4)$ and the shifting of the canonical center can be performed via a QR-decomposition scaling as $\mathcal{O}(\chi^3)$. In all applications of this method, we take the first product MPS as our candidate $\mat{A} = \mat{B}$ and sweep through the sites $c=1 \to c=n \to c=1$ two times.

Alternatives would be the zip-up algorithm proposed by Stoudenmire et al.~\cite{Stoudenmire_mult_2010}, which also scales in the bond dimension as $\mathcal{O}(\chi^4)$, or the algorithm proposed by Michaelidis et al.~\cite{michailidis_tensor_2024} with an improved asymptotic scaling of $\mathcal{O}(\chi^3)$. We chose the iterative method over the alternatives mainly because of the intention to perform MPS-LBM simulations on GPU. Using QR-decomposition, the iterative method can be implemented without SVD, which has been shown to be preferable for MPS computations on GPU~\cite{unfried_qr-svd_2023}. Moreover, both alternative algorithms rely on intermediate results of preferably flexible bond dimensions, which we found to significantly impede performance gains through JAX~\cite{jax2018github} just-in-time compilation, from which we expect further computational performance gains.

\subsection{Vector Fields in Scale-Ordered Matrix Product State Representation}
\begin{figure}[tb]
    \centering
    \includegraphics[width=0.7\linewidth]{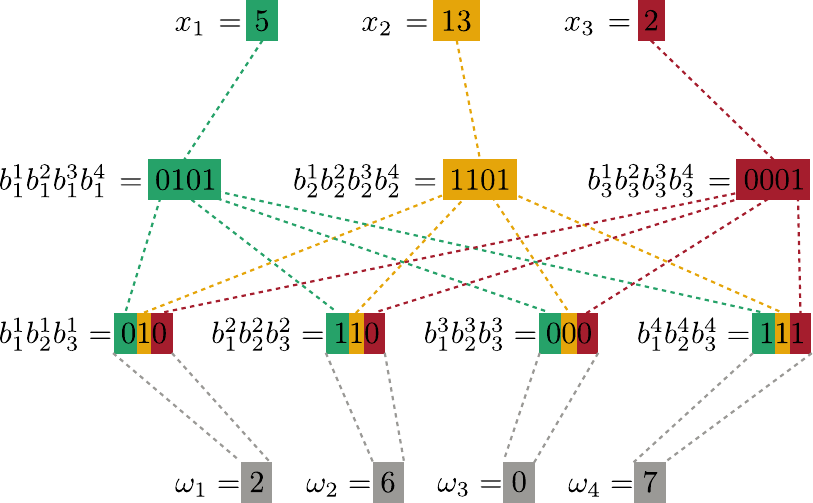}
    \caption{\edit{\textbf{Example of scale ordering of indices.} Exemplary reordering of indices with $D=3$ and $n=4$. The integer indices $x_i$ are represented in binary. The binary digits are regrouped in terms of their value. The resulting binary numbers are reinterpreted in terms of integers.}}
    \label{fig:scale_order}
\end{figure}

For numerical analysis, the scalar fields encountered in continuum mechanics are commonly discretized on a uniform grid within a computational domain. For example, consider the scalar field $f(\vec{x})$ discretized on a uniform grid within a computational domain of spatial dimension $D$. Coordinates are given as integer-valued indices $x_d$, such that $\sum_d x_d \vec{e}_d$ corresponds to the respective continuous coordinate, where $\vec{e}_d$ are the basis vectors of the lattice. For simplicity let the number of grid points in each spatial dimension be constrained to $2^n$, implying each discretized scalar field to be a tensor $\mat{f}^{x_1,\dots,x_D}$ of shape $\left( 2^n \right)^{\times D}$. Commonly $D=3$, thus naively decomposing $\mat{f}$ into an MPS would result in only three cores, limiting the possible compression from truncation. However, $n$ needs to be large to resolve small details in the field. Following the scale-ordering scheme introduced in \cite{gourianov_quantum-inspired_2022}, consider a binary representation of each spatial index as $x_i = b_i^1 \dots b_i^n$. Reordering the binary indices as
\begin{equation}
    \begin{aligned}
        &x_1, \dots, x_D = (b_1^1 \dots b_1^n), \dots, (b_D^1 \dots b_D^n)  \\
        \xrightarrow{\text{scale-ordering}} \quad &(b_1^1 \dots b_D^1), \dots, (b_1^n \dots b_D^n) = \omega_1, \dots, \omega_n,
    \end{aligned}
    \label{eq:scale_ordering}
\end{equation}
allows to reshape $\mat{f}$ into an order-$n$ tensor
\begin{equation}
    \mat{f}^{x_1,\dots, x_D} \xrightarrow{\text{scale-ordering}} \mat{f}^{\omega_1,\dots, \omega_n},
\end{equation}
which is well suited to MPS-decomposition as in \cref{eq:MPS}. \edit{See \cref{fig:scale_order} for an example with $D=3$ and $n=4$.}

\subsection{Matrix Product Operators for Shifting}\label{sec:mpo}
For the streaming step in LBM, we formulate shift MPO, i.e. linear operators $\mat{S_d}$ that shift all entries of a discretized scalar field by one grid node in a given direction $d$. 
\begin{equation}
    \mat{f}^{x_1,\dots,x_d,\dots, x_D} \xrightarrow{\mat{S_d}} \mat{f}^{x_1,\dots,x_{d+1},\dots, x_D}.
    \label{eq:shift}
\end{equation}
We construct the shift MPO for cyclic and non-cyclic shifts, and show that the MPO are of low bond dimension. The shift MPO is based on earlier work~\cite{gourianov_quantum-inspired_2022, holscher_quantum-inspired_2025}, and most of this derivation closely follows that of Kazeev et al.~\cite{kazeev_low-rank_2012}.

Consider for $D=2^n$ cyclic $\mat{S}^{(n)}$ and non-cyclic $\hat{\mat{S}}^{(n)}$ shift matrices. For $n=2$, these matrices are defined as
\begin{equation}
        \mat{S}^{(2)} =
        \begin{pmatrix}
            0 & 1 & 0 & 0\\
            0 & 0 & 1 & 0\\
            0 & 0 & 0 & 1\\
            1 & 0 & 0 & 0
        \end{pmatrix},
        \quad
        \hat{\mat{S}}^{(2)} =
        \begin{pmatrix}
            0 & 1 & 0 & 0\\
            0 & 0 & 1 & 0\\
            0 & 0 & 0 & 1\\
            0 & 0 & 0 & 0
        \end{pmatrix}.
\end{equation}
Using the $2\times2$ matrices
\begin{equation}
    {I} = \begin{pmatrix}
        1 & 0 \\
        0 & 1 \\
    \end{pmatrix}, \quad
    {J} = \begin{pmatrix}
        0 & 1 \\
        0 & 0 \\
    \end{pmatrix}, \quad
    {J}' = \begin{pmatrix}
        0 & 0 \\
        1 & 0 \\
    \end{pmatrix}, \quad
    {P} = \begin{pmatrix}
        0 & 1 \\
        1 & 0 \\
    \end{pmatrix},
\end{equation}
the shift matrices can be constructed as
\begin{equation}
    \begin{aligned}
        \mat{S}^{(2)} &= 
        \begin{pmatrix}
            {J} & {J}' \\
            {J}' & {J} 
        \end{pmatrix} = {I} \otimes {J} + {P} \otimes {J'}, \\
        \hat{\mat{S}}^{(2)} &= 
        \begin{pmatrix}
            J & J' \\
            0 & J
        \end{pmatrix} = {I} \otimes {J} + {J} \otimes {J'}.
    \end{aligned}
\end{equation}
For $n=3$, the shift matrices can be expressed in terms of $\mat{S}^{(2)}$ and $\hat{\mat{S}}^{(2)}$ as
\begin{equation}
    \begin{aligned}
        \mat{S}^{(3)} &= 
        \begin{pmatrix}
            \hat{\mat{S}}^{(2)} & J'^{\otimes2} \\
            J'^{\otimes2} & \hat{\mat{S}}^{(2)} 
        \end{pmatrix} = {I} \otimes \hat{\mat{S}}^{(2)} + {P} \otimes J'^{\otimes2}, \\
        \hat{\mat{S}}^{(3)} &= 
        \begin{pmatrix}
            \hat{\mat{S}}^{(2)} & J'^{\otimes2} \\
            0 & \hat{\mat{S}}^{(2)}
        \end{pmatrix} = {I} \otimes \hat{\mat{S}}^{(2)} + {J} \otimes J'^{\otimes2},
    \end{aligned}
\end{equation}
leading to the recursive definition
\begin{equation}
    \begin{aligned}
        \mat{S}^{(n)} &= 
        \begin{pmatrix}
            \hat{\mat{S}}^{(n-1)} & J'^{\otimes(n-1)} \\
            J'^{\otimes(n-1)} & \hat{\mat{S}}^{(n-1)} 
        \end{pmatrix} = {I} \otimes \hat{\mat{S}}^{(n-1)} + {P} \otimes J'^{\otimes(n-1)}, \\
        \hat{\mat{S}}^{(n)} &= 
        \begin{pmatrix}
            \hat{\mat{S}}^{(n-1)} & J'^{\otimes(n-1)} \\
            0 & \hat{\mat{S}}^{(n-1)}
        \end{pmatrix} = {I} \otimes \hat{\mat{S}}^{(n-1)} + {J} \otimes J'^{\otimes(n-1)}.
    \end{aligned}
    \label{eq:recursive}
\end{equation}
We introduce the rank core product as defined in~\cite{kazeev_low-rank_2012}. Consider 4-dimensional tensors $\mat{A}_{\alpha\beta}^{ab}$ and $\mat{B}_{\alpha\beta}^{ab}$, and explicitly expand the upper indices as
\begin{equation}
    \begin{bmatrix}
        \mat{A}_{11} & \mat{A}_{12} & \dots \\
        \mat{A}_{21} & \mat{A}_{22} & \dots \\
        \vdots & \vdots & \ddots \\
    \end{bmatrix}, \quad \begin{bmatrix}
        \mat{B}_{11} & \mat{B}_{12} & \dots \\
        \mat{B}_{21} & \mat{B}_{22} & \dots \\
        \vdots & \vdots & \ddots \\
    \end{bmatrix},
\end{equation}
where each entry is a 2-dimensional tensor.
The rank core product "$\Join$" of these two tensors is
\begin{equation}
    \begin{bmatrix}
    \mat{A}_{11} & \mat{A}_{12}\\
    \mat{A}_{21} & \mat{A}_{22}
    \end{bmatrix}
    \Join
    \begin{bmatrix}
    \mat{B}_{11} & \mat{B}_{12}\\
    \mat{B}_{21} & \mat{B}_{22}
    \end{bmatrix}
    =
    \begin{bmatrix}
    \mat{A}_{11} \otimes \mat{B}_{11} + \mat{A}_{12} \otimes \mat{B}_{21} &
    \mat{A}_{11} \otimes \mat{B}_{12} + \mat{A}_{12} \otimes \mat{B}_{22} \\[4pt]
    \mat{A}_{21} \otimes \mat{B}_{11} + \mat{A}_{22} \otimes \mat{B}_{21} &
    \mat{A}_{21} \otimes \mat{B}_{12} + \mat{A}_{22} \otimes \mat{B}_{22}
    \end{bmatrix}.
\end{equation}
For connecting this to MPO, consider a 3-dimensional tensor $\mat{T} \in \mathbb{R}^3$ and a linear operator
\begin{equation}
\begin{aligned}
    \mat{O}:\mathbb{R}^3 &\to \mathbb{R}^3 \\
    \mat{T}^{abc} &\mapsto  \sum_{a,b,c} \mat{O}^{def,abc} \ \mat{T}^{abc}.
\end{aligned}
\end{equation}
We define the MPO decomposition of $\mat{O}$ as
\begin{equation}
    \mat{O}^{abc,def} = \sum_{\alpha, \beta} \mat{A}_{\alpha}^{ad} \mat{B}_{\alpha\beta}^{be} \mat{C}_{\beta}^{cf},
\end{equation}
which can be written in terms of the rank core product as
\begin{equation}
    \mat{O} = \mat{A} \Join \mat{B} \Join \mat{C}.
\end{equation}
Similarly, we decompose $\mat{S}$ and $\hat{\mat{S}}$ with the rank core product and rewrite \cref{eq:recursive} as
\begin{equation}
\begin{aligned}
        \mat{S}^{(n)} &=
        \begin{bmatrix}
            I & P
        \end{bmatrix}
        \Join
        \begin{bmatrix}
            \hat{\mat{S}}^{(n-1)} \\
            J'^{\otimes(n-1)}
        \end{bmatrix}, \\
        \hat{\mat{S}}^{(n)} &=
        \begin{bmatrix}
            I & J
        \end{bmatrix}
        \Join
        \begin{bmatrix}
            \hat{\mat{S}}^{(n-1)} \\
            J'^{\otimes(n-1)}
        \end{bmatrix},
\end{aligned}
\end{equation}
which indicates that only the first core differs for cyclic and non-cyclic shift MPO. $\hat{\mat{S}}$ can be further decomposed as
\begin{equation}
\begin{aligned}
    \hat{\mat{S}}^{(n)} &=
        \begin{bmatrix}
            I & J
        \end{bmatrix}
        \Join
        \begin{bmatrix}
            I & J \\
            0 & J'
        \end{bmatrix}
        \Join
        \begin{bmatrix}
            \hat{\mat{S}}^{(n-2)} \\
            J'^{\otimes(n-2)}
        \end{bmatrix} \\
        &=
        \begin{bmatrix}
            I & J
        \end{bmatrix}
        \Join
        \underbrace{
        \begin{bmatrix}
            I & J \\
            0 & J'
        \end{bmatrix}
        \Join
        \dots
        \Join
        \begin{bmatrix}
            I & J \\
            0 & J'
        \end{bmatrix}
        }_{(n-2)-\text{times}}
        \Join
        \begin{bmatrix}
            \hat{\mat{S}}^{(1)} \\
            J'
        \end{bmatrix}. \\
\end{aligned}
\end{equation}
Using $\hat{\mat{S}}^{(1)} = J$ we can define 
\begin{equation}
    \mat{A} = \begin{bmatrix}
        I & P
    \end{bmatrix}, \quad
    \hat{\mat{A}} = \begin{bmatrix}
        I & J
    \end{bmatrix}, \quad
    \mat{B} = \begin{bmatrix}
        I & J \\
        0 & J'
    \end{bmatrix}, \quad
    \mat{C} = \begin{bmatrix}
        J \\
        J'
    \end{bmatrix},
\end{equation}
such that
\begin{equation}
    \begin{aligned}
        \mat{S}^{(n)} &= \mat{A} \Join \mat{B} \Join \dots \Join \mat{B} \Join \mat{C}, \\
        \hat{\mat{S}}^{(n)} &= \hat{\mat{A}} \Join \mat{B}\Join \dots \Join \mat{B} \Join \mat{C}.
    \end{aligned}
\end{equation}
It is straightforward to show that a shift in the opposite direction is achieved by simply interchanging $J$ with $J'$ and vice versa in all cores.

The above derivation only produces shift matrices on vectors. However, for MPS-LBM, we need to shift in all physical dimensions separately. Note that in \cref{eq:scale_ordering}, the indices are split into binary digits such that the $i$-th bit of each core corresponds to the $i$-th dimension. Thus, the cores of the shift MPO are to act on the $i$-th index only to result in a shift in the $i$-th dimension. This is achieved by padding the individual cores with identities. Say we want a shift in the first of three dimensions, then the first core of the respective MPO is
\begin{equation}
    A\Join\begin{bmatrix}
        I & 0 \\
        0 & I
    \end{bmatrix}
    \Join
    \begin{bmatrix}
        I & 0\\
        0 & I
    \end{bmatrix},
\end{equation}
and analogously $\mat{B}$ and $\mat{C}$ are padded.

\section{Matrix Product States in the Lattice Boltzmann Method}\label{sec:mpslbm}

\subsection{Lattice Boltzmann Method}
The lattice Boltzmann method (LBM) is a mesoscopic model of fluid dynamics, approximating the evolution of macroscopic flow governed by the Navier-Stokes equations through discretized \emph\edit{particle distribution functions} $f_i$. 
The evolution of single components of \edit{particle distribution functions} is governed by the lattice Boltzmann equation~\cite{chen_recovery_1992,chen_lattice_1998} 
\begin{equation} 
    f_i(\vec{x}+\vec{c}_i \Delta t, t + \Delta t) = f_i(\vec{x}, t) + \Omega_i(\vec{x}, t), 
    \label{eq:evolution} 
\end{equation} 
with $\Omega_i$ representing the collision operator. 
This process consists of two steps: (i) \emph{collision}, a local relaxation modeling local particle collision by the redistribution of particles among distribution functions $f_i$ (given by the RHS of \cref{eq:evolution}), and (ii) \emph{streaming}, which propagates \edit{particle distribution functions} along the lattice (shifting of space and time coordinates on the LHS of \cref{eq:evolution}).
For simplicity, we employ the BGK collision operator~\cite{bhatnagar_model_1954} which involves a linear relaxation with rate $\tau$ towards an equilibrium $f^{eq}_i$
\begin{equation}
    \Omega_i(\vec{x}, t) = \frac{\Delta t}{\tau} \left( f^{eq}_{i}(\vec{x}, t)-f_{i}(\vec{x}, t) \right),
    \label{eq:collision}
\end{equation}
 with the \emph{equilibrium distribution} $f^{eq}$ defined as 
\begin{equation}
    f^{eq}_{i}(\vec{x}, t) = w_i\rho\left(1 + \frac{\vec{u} \cdot \vec{c}_i}{c_s^2} + \frac{(\vec{u} \cdot \vec{c}_i)^2}{2c_s^4} - \frac{\vec{u} \cdot \vec{u}}{2c_s^2} \right).
    \label{eq:equilibrium}
\end{equation}
The \emph{relaxation time} $\tau$ is determined by the kinematic viscosity $\nu$ via $\tau = \frac{\Delta t}{2} + \frac{\nu}{c_s^2}$.
Macroscopic quantities such as density $\rho$ and momentum $\rho \vec{u}$ are recovered from the hydrodynamic moments~\cite{abe_derivation_1997,he_theory_1997} of these \edit{particle distribution functions} 
\begin{equation} 
    \rho = \sum_{i=1}^Q f_i(\vec{x}, t), \qquad \rho \vec{u} = \sum_{i=1}^Q \vec{c}_i f_i(\vec{x}, t), 
    \label{eq:macro} 
\end{equation} 
where $Q = |\{\vec{c}_i\}|$ denotes the number of discrete microscopic velocities $\vec{c}_i$. 

\subsection{MPS Lattice Boltzmann Method}
The lattice Boltzmann method based on the MPS framework (MPS-LBM) leverages the mostly local formulation of LBM. 
Distribution functions $f_i$ are spatially discretized on uniform Cartesian meshes. An MPS-decomposition of spatial dimensions enables compression, reducing the number of variables parameterizing the state with controllable accuracy.

All considered scalar fields, i.e, the \edit{particle distribution functions} $f_i$ and macroscopic variables $\vec{u}$ and $\rho$, are discretized on a uniform grid within a computational domain of spatial dimension $D$. Following \cref{eq:scale_ordering} they are brought into scale-ordered form to enable effective MPS decomposition.
Local collision interactions consist mostly of element-wise addition and multiplication, while the streaming step admits an exact formulation as a low-rank matrix product operator. As shown in \cref{sec:mps}, these operations can be performed efficiently without leaving the compressed MPS manifold. The only operation that cannot be directly handled by the MPS formalism is the computation of velocity $u$ via the first moment of the \edit{particle distribution functions} in \cref{eq:macro}. We approximate the normalization factor $1/\rho$ to circumvent the lack of a closed-form algorithm for elementwise inversion of an MPS. Sanavio and Succi~\cite{sanavio_carleman_2024} proposed a Taylor expansion of this term around the constant value $1$ for a Carleman linearization algorithm of the LBM, which is a suitable approximation for weakly compressible single-phase flows. 
We build upon this idea by introducing a second-order Taylor expansion of $1/{\rho}$ around the global mean density $\rho_0$ at each time step $t$
\begin{equation}
    \frac{1}{\rho}\approx \frac{1}{\rho_0}-\frac{\delta\rho}{\rho_0^2}+\frac{(\delta\rho)^2}{\rho_0^3}+\mathcal{O}\left((\delta\rho)^3\right),
\end{equation}
where $\delta\rho=\rho-\rho_0$ denotes the density fluctuation. 
The mean density $\rho_0$ is computed by contracting the $\rho$-MPS with a rank-1 unit MPS and normalizing the result. 
For weakly compressible flows, the relative density fluctuation scales with the Mach number $\text{Ma}$ as $ \frac{\delta\rho}{\rho_0} \sim \mathcal{O}(\text{Ma}^2) $ in the limit $ \text{Ma} \to 0 $~\cite{he1997lattice}. Consequently, the second-order approximation of $ 1/{\rho} $ introduces an error that scales as $ \mathcal{O}(\text{Ma}^6) $. The introduced error is several magnitudes smaller than the inherent compressibility error $ \mathcal{O}(\text{Ma}^2) $ of the LBM~\cite{kruger_lattice_2017}. This approach is validated by numerical experiments presented in \ref{sup:inverse}.
\edit{While this approximation is robust within the weakly compressible regime, its accuracy deteriorates outside of it. For simulations involving large density variations, e.g. multiphase flows with steep density gradients or fully compressible flows, the condition $\delta\rho/\rho_0 \ll 1$ no longer applies, rendering the low-order Taylor expansion inaccurate.}

The runtime of MPS-LBM is dominated by elementwise addition, multiplication, and the application of the low-rank shift MPO.
Addition of two MPS is performed as in \cref{eq:add}, which yields bond dimensions on the order of $2 \chi$. Reducing the bond dimension via an iteration of singular value decompositions (SVD) introduces a complexity scaling of $\mathcal{O}\left(n\chi^3\right)$ for this operation, where $n$ is the length of the MPS determined by the grid resolution. 
As shown in \cref{sec:mpo}, the bond dimension of shift MPO is $\chi\leq3$, and hence the application of shift MPO to MPS followed by an iteration of SVD also results in a scaling of $\mathcal{O}\left(n\chi^3\right)$.
Following our discussion in \cref{sec:mul}, elementwise multiplication of two MPS scales as $\mathcal{O}\left(n\chi^4\right)$. In total, we find that our algorithm is dominated by elementwise multiplication and thus exhibits an overall cost scaling of $\mathcal{O}\left(n\chi^4\right)$ matching the asymptotic runtime of previous quantum-inspired approaches to CFD. Note that any of the aforementioned primitives can be straightforwardly substituted with further improvements in terms of MPS operations and their implementation in high-performance libraries. We provide the results of numerical runtime experiments on current GPU hardware in \ref{sup:runtime}.

\subsubsection{Masking of Objects and Boundaries}
In lattice Boltzmann simulations, the computation of immersed objects or non-periodic boundary conditions requires localized computations that only affect cells on the edge of the domain or cells at the boundary of an object. 
In conventional implementations, these nodes are readily accessible via explicit indexing. 
However, the inherently non-local and compressed structure of MPS representations prevents direct access to individual boundary cells.
In MPS-LBM, a combination of non-cyclic shifts and low-rank binary MPS marking boundary cells is employed to enable boundary-specific computations. 

\paragraph{Non-Periodic Boundaries}
Non-periodic boundaries are treated by replacing the cyclic shift matrices ${\mat{S}}$ used in the streaming step with non-cyclic ones $\hat{\mat{S}}$ at the corresponding boundaries. Upon shifting, this introduces zero rows on the set of boundary cells $\mathcal{B}$ corresponding to particle distributions originating from outside of the computational domain. Inside the domain, $\hat{\mat{S}}$ continues to perform shifts consistent with the streaming operation. We employ a binary mask $m_{\mathcal{B}}^{\vec{x}}$, that is $1$ for all $x \in \mathcal{B}$, and $0$ elsewhere. If $x_i=0$ ($x_i=N$) is a periodic boundary and $a_{i,d}=1$ ($a_{i,d}=-1$) for some $d$, then modified streaming
\begin{equation}
    f_i^{\vec{x}} \mapsto \left( \bigotimes_{d=1}^D \mat{S}({a_{i,d}}) \right) f_i^{\vec{x}}
    \label{eq:boundaries1}
\end{equation}
is applied. Otherwise we apply
\begin{equation}
    f_i^{\vec{x}} \mapsto m_{\mathcal{B}}^{\vec{x}}\ b(f_{\hat{i}}^{\vec{x}}, \rho_b, \vec{u}_b) + \left( \bigotimes_{d=1}^D \hat{\mat{S}}({a_{i,d}}) \right) f_i^{\vec{x}}.
    \label{eq:boundaries2}
\end{equation}
Here, $b(f_{\hat{i}}^{\vec{x}}, \rho_b, \vec{u}_b)$ denotes a general boundary function that computes the appropriate values based on the incoming distribution $f_{\hat{i}}^{\vec{x}}$, boundary density $\rho_b$, and velocity $\vec{u}_b$.
For a no-slip boundary~\cite{he_analytic_1997,ziegler_boundary_1993} the function $b_{\text{no-slip}}$ is given by the bounce-back rule as
\begin{equation}
    b_{\text{no-slip}}(f_{\hat{i}}^{\vec{x}}, \rho_b, \vec{u}_b) = f_{\hat{i}}^{\vec{x}}.
    \label{eq:no_slip}
\end{equation}
Dirichlet boundaries with prescribed velocity $\vec{u}_b$ are given by
\begin{equation}
    b_{\text{velocity}}(f_{\hat{i}}^{\vec{x}}, \rho_b, \vec{u}_b) = f_{\hat{i}}^{\vec{x}} - \frac{2w_i \rho_b}{c_s^2} \vec{c}_i \cdot \vec{u}_b,
\end{equation}
where $\rho_b$ is the density at the boundary. We set $\rho_b=\rho_0$, assuming weak compressibility in the computational domain. 
Outlet boundaries~\cite{izquierdo_characteristic_2008} are computed by describing a wall density $\rho_b$ and using the velocity $\vec{u}(\vec{x}\in \mathcal{B}, t) = \vec{u}_b$ 
\begin{equation}
    b_{\text{pressure}}(f_{\hat{i}}^{\vec{x}}, \rho_b, \vec{u}_b) = -f_{\hat{i}}^{\vec{x}} + 2w_i \rho_b \left( 1+ \frac{(\vec{c}_i \cdot \vec{u}_b)^2}{2c^4_s} -\frac{\vec{u}_b \cdot \vec{u}_b}{2c_s^2} \right).
\end{equation}

\paragraph{Immersed Objects}
For immersed objects, we apply the same no-slip condition as defined in \cref{eq:no_slip}. However, unlike the approach used for non-cyclic boundaries in \cref{eq:boundaries2}, we do not modify the streaming step directly. Instead, we allow the streaming to proceed as if no obstacles were present, and subsequently reverse the streaming for the solid internal boundary cells. This reversal is conditioned by a binary MPS mask $m_{\mathcal{O}}^{\vec{x}}$, which takes the value $1$ at all cells $x \in \mathcal{O}$ inside the object and $0$ elsewhere.

The implementation is
\begin{equation}
    f_i^{\vec{x}} \mapsto (1-m_{\mathcal{O}}^{\vec{x}}) \ f_i^{\vec{x}} + \left( \bigotimes_{d=1}^D \mat{S}({a_{i,d}}) \right) m_{\mathcal{O}}^{\vec{x}} \ f_{\hat{i}}^{\vec{x}}.
    \label{eq:objects}
\end{equation}

While the multiplication with $m_\mathcal{O}^{\vec{x}}$ ensures bounce-back, $1-m_\mathcal{O}^{\vec{x}}$ deletes the remaining non-zero entries inside the solid object.

\section{Experimental Setup}\label{sec:setup}
\subsection{Test Settings}\label{sec:setting}
In LBM, a unit conversion from \emph{physical} to \emph{lattice} units is applied. For all experiments, this conversion is given by the velocity conversion factor $\Delta u = \frac{u_\text{physical}}{u_\text{lattice}} = \frac{u_0}{\text{Ma} \ c_s}$. Length is discretized using $\Delta x = \frac{L}{N}$ and the time step is determined by $\Delta t = \frac{\Delta x}{\Delta u}$. Here, $u_0$ is the characteristic velocity of the flow problem, $L$ is the side length of the domain, and $N$ is the number of lattice sites per side. Two-dimensional and three-dimensional simulations use the D2Q9 and D3Q15 lattice schemes, respectively, which define the lattice speed of sound as  $c_s = 1/\sqrt{3}$.

The three-dimensional Taylor-Green vortex (TGV) is defined on a cube of edge length $L=2\pi$ and a characteristic velocity $u_0=1$ with initial conditions
\begin{equation}
    \begin{aligned}
        u(x,y,z) &= -  \sin\left(  x\right) \cos\left(  y\right) \cos\left( z\right),\\
        v(x,y,z) &=  \cos\left(  x\right) \sin\left(  y\right) \cos\left(  z\right),\\
        w(x,y,z) &= 0,\\
        \rho(x,y,z) &= 1 + \frac{1}{16\Delta u^2} \left(\cos\left( 2 x \right) + \cos\left( 2 y \right)\right)\cos\left( 2 z\right).
    \end{aligned}
\end{equation}
In accordance with similar experiments in~\cite{nathen_lbm-tgv_2018,gourianov_quantum-inspired_2022}, we define the Reynolds number as $\text{Re} = \frac{u_0L}{2\pi\nu}$. All 3D TGV simulations were conducted with $\text{Ma}=0.1/c_s$ at a resolution of $N^3 = 256^3$.

\begin{figure}[htb]
    \centering
    \includegraphics[width=.95\linewidth]{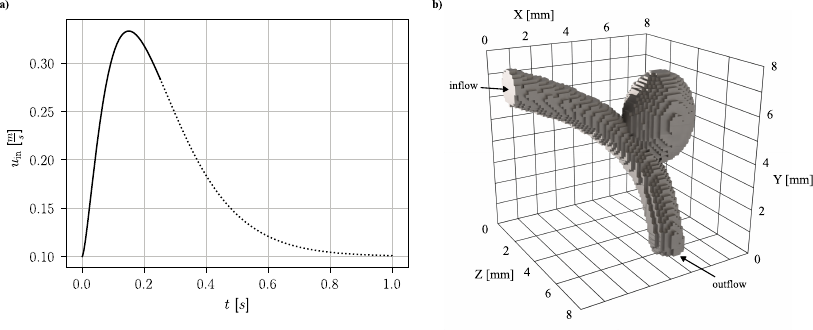}
    \caption{\textbf{Setup of the aneurysm simulation}. \textbf{a)} Time-dependent magnitude of the flow velocity at the inflow. The simulations were performed until $t=0.25s$ depicted by the solid line. For reference, the dotted line shows the remaining part of the $1s$ heartbeat cycle. \textbf{b)} Full geometry as used in the aneurysm test case. The time-dependent inflow is at $z=0$, and the outflow boundary is at $y=0$.}
    \label{fig:an_setup}
\end{figure}

To ensure realistic flow conditions within the aneurysm geometry, we base our simulation parameters on those reported by Horvat et al.~\cite{horvat_aneurysm_2025}, with appropriate adjustments to account for the specifics of our computational setup. The velocities averaged over the inflow surface, ranging between $0.1~\si{m\per s}$ and $0.34~\si{m\per s}$, are approximated by $v_\text{in}(t) = 0.1 + 18t^{1.5}e^{-10t}~\si{m\per s}$ (see \subfigref{fig:an_setup}{a}). The domain is a cube of edge length $8~\si{mm}$ discretized at a resolution of $N^3=64^3$. The idealized aneurysm geometry is shown in \subfigref{fig:an_setup}{b}. The aneurysm has a size of approximately $l_\text{an}\approx4\si{mm}$. The relaxation time is given by $\tau = \frac{\text{Ma}N\nu}{Lu_0c_s}$, where $u_0 \approx0.78~\si{m\per s}$ is the maximum velocity observed at the center of the blood vessel during simulation. The viscosity was adjusted to $\nu=1.12\times10^{-5}~\si{m^2 \per s}$ to ensure numerical stability of the computational setup. This results in a Reynolds number of $\text{Re}=\frac{l_\text{an}u_0}{\nu}\approx280$. As an initial condition for the \edit{particle distribution functions}, a stationary flow was precomputed using a constant inflow velocity of $0.1~\si{m \per s}$ over a duration of $0.15~\si{s}$.

\begin{figure}[htb]
    \centering
    \includegraphics[width=.95\linewidth]{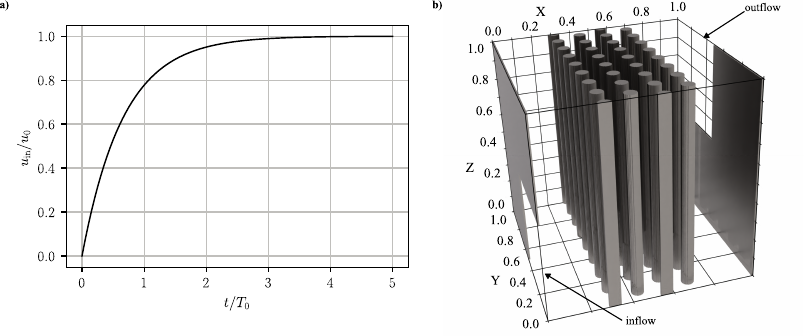}
    \caption{\textbf{Setup of the pin-fin heat sink simulation}. \textbf{a)} Time-dependent magnitude of the flow velocity at the inflow. \textbf{b)} Full geometry as used in the pin-fin test case. The time-dependent inflow is at $x=0$, and the outflow boundary is at $x=L$.}
    \label{fig:pf_setup}
\end{figure}

In developing the parameterization of the pin-fin heat sink, we follow~\cite{Shahsavar_pinfin_2021}. To ensure applicability across a range of geometric length scales, the heat-sink configuration is defined on a normalized cubic domain of edge length $L$ (see \subfigref{fig:pf_setup}{b}). The simulation is initiated with a vanishing velocity field, and the inflow velocity is ramped up as $u_\text{in}(t) = u_0\left( 1-\exp(-1.5t) \right)$ (see \subfigref{fig:pf_setup}{a}). The Reynolds number is $\mathrm{Re} =\frac{3u_0L}{16\nu}= 400$ with respect to the distance between the layers of pins $l_\textrm{layer} = L/8$ and the maximum velocities occurring towards the end of the simulation of around $u_\mathrm{max}\approx\frac{3}{2}u_0$. The simulations were run up to $t/T_0=5$ with $T_0=L/u_0$.

\subsection{Quantitative Metrics}\label{sec:metrics}
We evaluate the MPS-LBM accuracy using the relative error of the macroscopic velocity field $\vec{u}$ in the $\text{l}^2$-norm given by
\begin{equation}
    \epsilon_{\text{l}^2} = \frac{\sqrt{\sum_{d,\vec{x}} \left(u_{d,\vec{x}}-\hat{u}_{d,\vec{x}}\right)^2}}{\sqrt{\sum_{d,\vec{x}} \hat{u}^2_{d,\vec{x}}}},
\end{equation}
where $u_{d,\vec{x}}$ is the result of MPS-LBM, index $d$ is the spatial dimension of the velocity field at position $\vec{x}$ and $\vec{\hat{u}_{d,\vec{x}}}$ is the reference result.

The total kinetic energy $E_\text{kin}$ for TGV is given by
\begin{equation}
    E_\text{kin} = \frac{\rho V_c}{2}\sum_d u_{d,\vec{x}}^2,
\end{equation}
where $V_c$ is the volume per lattice site. Density is $\rho = 1/L^3$ such that the total mass is normalized to $1$. The total energy dissipation $-\frac{d}{dt} E_\text{kin}$ is numerically evaluated using second-order finite differences.

Pressures $p$ are normalized in terms of the prescribed pressure at the outflow $p_0$ and averaged over the unmasked parts of the region of interest $\mathcal{D}$
\begin{equation}
    \Delta p/p_0 = \frac{1}{p_0\lvert \mathcal{D} \rvert} \sum_{\vec{x}\in\mathcal{D}} \left(p_{\vec{x}}-p_0\right).
\end{equation}
In the aneurysm case, we average over the region around the aneurysm $0.53L<x<0.97L, 0.39L<y<0.93L, 0.33L<z<0.69L$. In the pin-fin configuration, we are interested in the pressure drop from inflow to outflow, so we average over the inflow region $0<x<2L/256, 0<y<L/2, 0<z<L/2$.

The compression ratio CR is
\begin{equation}
    \text{CR} = \frac{\text{NVPS}_\text{ref}}{\text{NVPS}},
\end{equation}
where the number of variables parametrising the solution (NVPS) of a component of the standard LBM on a $(2^n)^D$ grid is $\text{NVPS}_\text{LBM}=(2^n)^D$. For MPS-LBM, it is given as
\begin{equation}
    \text{NVPS}_\text{MPS} = \sum_{i=1}^n \min\left(\left(2^D\right)^{i-1}, \left(2^D\right)^{n-i+1},\chi\right) \times 2^D \times \min\left(\left(2^D\right)^{i}, \left(2^D\right)^{n-i},\chi\right).
\end{equation}

\section{Results}\label{sec:results}

\subsection{Three Dimensional Taylor-Green Vortex}\label{sec:3dtgv}
\begin{figure}[htb]
    \centering
    \includegraphics[width=\linewidth]{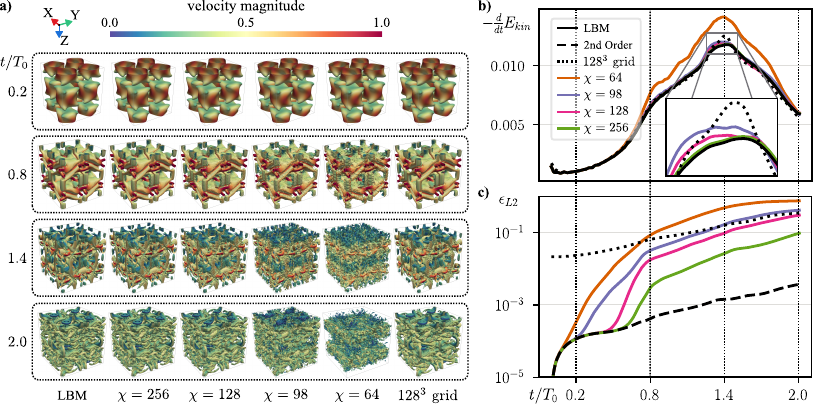}
    \caption{\textbf{Results of the 3D Taylor-Green vortex}. a) Vortical structures, identified by the Q-criterion isosurfaces, are shown at four distinct times $t/T_0 = 0.2,0.8,1.4,2$ for LBM on $256^3$ mesh, MPS-LBM with $\chi = 256, 128, 98, 64$, and a coarse LBM. \textbf{b)} Total kinetic energy dissipation $-\frac{d}{dt} E_\text{kin}$, colored solid lines depict MPS-LBM results, solid black line fine LBM, dotted black line coarse LBM. \textbf{c)} The relative $\text{l}^2$ error $\epsilon_{\text{l}^2}$ with respect to fine LBM. Solid colored lines represent MPS-LBM, dotted black corresponds to coarse LBM. For comparison, the values of fine LBM with second-order approximation of $1/\rho$ are given as the dashed black line.}
    \label{fig:tgv}
\end{figure}

We simulate the three-dimensional Taylor-Green vortex (TGV) at a Reynolds number of $\text{Re} = 800$ on a grid of size $256^3$. To investigate the impact of compression on simulation accuracy, we conduct experiments using MPS-LBM with different bond dimensions $\chi = \{256, 128, 98, 64\}$. Reference results are obtained using three configurations: (i) a fine LBM on a $256^3$ mesh, (ii) fine LBM with a second-order approximation of $1/\rho$ on $256^3$ mesh, and (iii) coarse LBM on a $128^3$ mesh.

\subfigref{fig:tgv}{a} depicts all simulation results at four different times $t/T_0 = 0.2, \allowbreak 0.8,\allowbreak 1.4,\allowbreak 2.0$, where $T_0 = L / u_0$. Here, $u_0$ denotes the maximum velocity magnitude in the initial conditions, and $L$ is the edge length of the domain.
All simulations capture the large-scale structures at time $t/T_0 = 0.2$ well. The solution trajectory deviates from the reference for $\chi = 64$ at $t/T_0 = 0.8$ and for $\chi = 98$ at $t/T_0 = 1.4$, as compression prohibits the formation of smaller-scale structures that emerge later in the flow. Simulations using $\chi = 128$ and $\chi = 256$ reproduce all relevant details throughout the simulation, with $\chi = 128$ showing only minor deviations from fine LBM. In comparison, the coarse LBM falls between the MPS-LBM results with $\chi = 256$ and with $\chi = 128$.

In \subfigref{fig:tgv}{b}, the energy dissipation $-\frac{d}{dt}E_\text{kin}$ is shown. All simulations qualitatively follow the fine LBM and produce physically consistent behavior. The simulation with $\chi = 64$ deviates significantly from the baseline, with the largest errors near the dissipation peak. All MPS-LBM simulations show slightly elevated dissipation before the peak at $t/T_0 \approx 1.3$, while the under-resolved $128^3$ LBM exhibits the highest dissipation, marked by a sharp peak at $t/T_0 = 1.4$.

\subfigref{fig:tgv}{c} shows the relative error in the $L^2$-norm $\epsilon_{\text{l}^2}$ of the macroscopic velocity field for all simulation cases compared to the reference LBM result. As an additional reference, the error of fine LBM with second-order approximation of $1/\rho$ is included. This comparison helps to distinguish the influence of MPS compression from that of the $1/\rho$ approximation.
At $t/T_0 = 0.2$, the curve for $\chi = 64$ already separates from the second-order approximation, while higher bond dimensions introduce no noticeable additional inaccuracies. This changes at $t/T_0 = 0.8$, where simulations with lower bond dimensions exhibit higher errors. From that point onward, all simulations show a gradual increase at error, with approximately half an order of magnitude separating $\chi = 64$, $\chi = 98$, and $\chi = 256$. The coarse LBM starts with a substantially higher initial error. However, it matches $\chi = 64$ at $t/T_0 = 0.8$ and $\chi = 98$ at $t/T_0 = 1.4$. The final error at $t/T_0 = 2$ is close to that of MPS-LBM with a bond dimension of $\chi = 128$.

For bond dimensions $\chi = 256$ and $\chi = 128$, the compression ratios are $\text{CR}_{256} \approx 13$ and $\text{CR}_{128} \approx 42$, respectively. A $128^3$ grid yields $\text{CR} \approx 8$, while $\chi = 98$ achieves $\text{CR}_{98} \approx 64$, equivalent to a $64^3$ grid. Lower grid resolutions are infeasible due to stability issues arising from a small relaxation time $\tau$. These results demonstrate that MPS-LBM outperforms simple grid-size reduction in both accuracy and compression efficiency.

\subsection{Aneurysm}
\begin{figure}
    \centering
    \includegraphics[width=\linewidth]{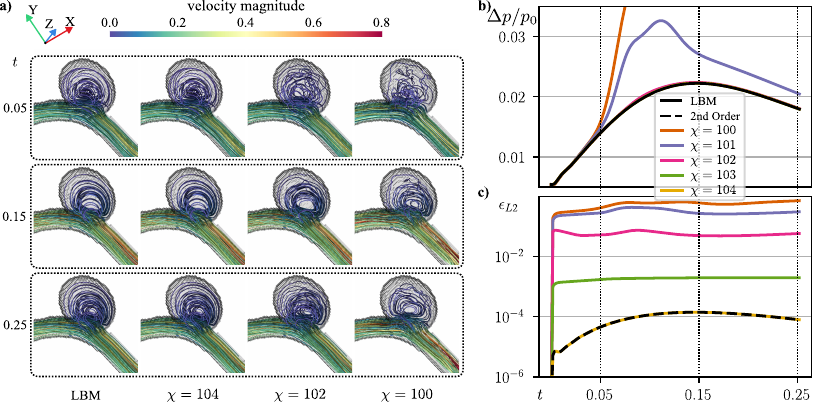}
    \caption{\textbf{Results for flow through a 3D aneurysm}. \textbf{a)} Streamlines of the flow at $t = 0.05,0.15,0.25$, for the fine LBM (first column) and MPS-LBM at $\chi = 104, 102, 100$ (columns 2-4). \textbf{b)} Normalized pressure difference $\Delta p/p_0$ averaged over the region of the aneurysm. \textbf{c)} Relative $\text{l}^2$ error $\epsilon_{\text{l}^2}$ with respect to the fine LBM. Solid colored lines represent MPS-LBM, the solid black line fine LBM, the dashed black line fine LBM with second-order approximation of $1/\rho$.}
    \label{fig:an}
\end{figure}

To demonstrate the versatility of MPS-LBM, we simulate blood flow through an aneurysm using time-dependent boundary conditions and a complex geometry encoded via a precomputed MPS boundary mask. Simulations were performed on a $64^3$ grid with bond dimensions $\chi = 104, 103, 102, 101, 100$ and compared to uncompressed LBM with exact $1/\rho$ and 2nd-order $1/\rho$ approximated. To isolate the effect of compressing the flow field on the overall accuracy, the mask bond dimension was fixed at $\chi_{\text{mask}} = 98$, which is sufficient to represent the full $64^3$ mask to machine precision. The simulation captures the first quarter of a one-second heartbeat, including peak inflow dynamics.

Streamline plots from the LBM simulation at $t = 0.05, 0.15, 0.25$ are shown in the first column of \subfigref{fig:an}{a}. At $t = 0.05$, the flow exhibits a weak vortical pattern with low velocities. By $t = 0.15$, near peak inflow, the vortex intensifies with higher velocities. At $t = 0.25$, the structure persists but weakens. Compared to this baseline, MPS-LBM at $\chi = 104$ yields nearly identical results.
Cases with bond dimensions of $\chi = 100$ and $\chi = 102$ exhibit distorted vortex structures within the aneurysm and fail to reproduce its detailed flow features.

As a relevant measure, we show the average pressure $\Delta p/p_0$ inside the aneurysm normalized with the prescribed outflow pressure in \subfigref{fig:an}{b}. The standard LBM shows a smooth increase of pressure towards the peak inflow at $t=0.15$ and a slight decrease afterwards. At $t=0.05$, the pressures obtained with MPS-LBM at $\chi \leq 101$ diverge from the baseline. At higher bond dimensions, MPS-LBM results are nearly indistinguishable from standard LBM. This suggests that the reproduced vortex structures at $\chi=102$ can maintain relevant physical aspects of the flow.

The $\text{l}^2$ error $\epsilon_{\text{l}^2}$ in \subfigref{fig:an}{c} supports these observations. For $\chi = 100$ and $101$, errors exceed $0.2$ and remain high throughout the simulation. At $\chi = 102$, errors drop by nearly an order of magnitude, and at $\chi = 103$ they stabilize around $10^{-3}$. With $\chi = 104$, MPS-LBM becomes indistinguishable from LBM with second-order $1/\rho$ approximation, indicating that MPS compression does not introduce significant additional error at this bond dimension.

Notably, the bond dimension $\chi$ of the MPS decomposed \edit{particle distribution function} must exceed $\chi_{\text{mask}}$ by only a small margin to achieve vanishing errors. However, as $\chi$ approaches $\chi_{\text{mask}}$, the simulation results quickly deviate strongly from the reference, suggesting that $\chi_{\text{mask}}$ serves as a lower bound for computationally meaningful $\chi$. At $\chi = 104$, a compression ratio of $\text{CR}_{104} \approx 2.3$ is achieved.

\subsection{Pin-Fin Configuration}
\begin{figure}
    \centering
    \includegraphics[width=\linewidth]{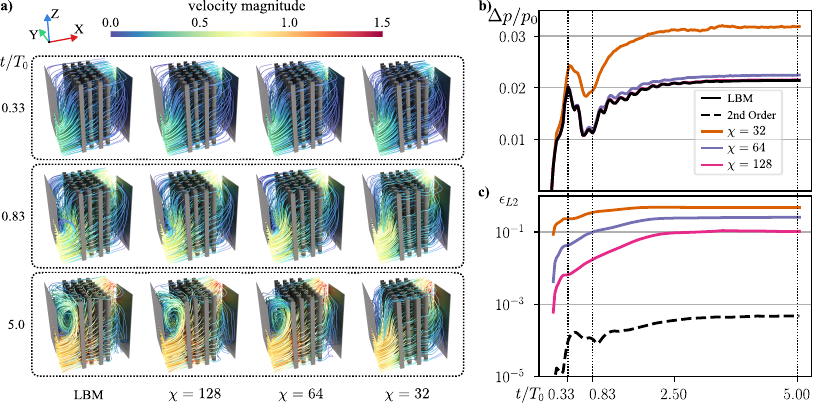}
    \caption{\textbf{Results for flow through a pin-fin configuration}. \textbf{a)} Streamlines of the flow at $t/T_0 =0.33,0.83,5.0$, for standard LBM (first column) and MPS-LBM at $\chi = 128, 64, 32$ (columns 2-4).  \textbf{b)} Normalized pressure difference $\Delta p/p_0$ averaged over the inflow region. \textbf{c)} Relative $\text{l}^2$-error $\epsilon_{\text{l}^2}$ with respect to the standard LBM. Solid colored lines represent MPS-LBM, solid black line standard LBM, the dashed black line standard LBM with second-order approximation of $1/\rho$.}
    \label{fig:pf}
\end{figure}

A pin-fin array, commonly used in industrial heat exchangers, exhibits translational symmetry and achieves high compression rates with MPS-LBM. Simulations were conducted on a $256^3$ grid. We compare MPS-LBM with bond dimensions $\chi = 128, 64, 32$ to standard LBM with exact and 2nd-order $1/\rho$ approximation. 
The mask bond dimension was set to $\chi_{\mathrm{mask}} = 16$ to ensure machine-precision accuracy.
Inflow velocity ramps from zero to $u_{\mathrm{in}} = 1$, and simulations run until $t/T_0 = 5$, where $T_0 = L/u_{\mathrm{in}}$, allowing the flow to become steady. The resulting vortex structures reveal that the flow becomes increasingly complex at higher Reynolds numbers.

The streamline plots in \subfigref{fig:pf}{a} depict the velocity field of uncompressed LBM and MPS-LBM simulations. At startup $t/T_0 = 0.33$, all simulations capture the flow accurately. By $t/T_0 = 0.83$, vortices begin to form above the inflow, which are resolved by simulations with $\chi \geq 64$. At the final time $t/T_0 = 5$, all cases except $\chi = 32$ reproduce detailed velocity structures, including the inflow vortex. The $\chi = 32$ simulation fails to capture these features.

The normalized pressure drop $\Delta p/p_0$ from inflow to outflow is shown in \subfigref{fig:pf}{b}. All simulations accurately predict the pressure drop up to $t/T_0 = 0.33$. Beyond this point, $\chi = 32$ overshoots, following the overall trend but overestimating even after convergence to the steady state. Other simulations track the reference closely, with $\chi = 64$ slightly overshooting and $\chi = 128$ matching the pressure drop exactly.

\subfigref{fig:pf}{c} shows a clear separation with respect to the examined bond dimensions in terms of the $\text{l}^2$-error $\epsilon_{\text{l}^2}$ of the velocity field. $\chi=32$ stabilizes around $0.5$, $\chi=64$ around $0.2$ and $\chi=128$ at $0.1$. The error of the standard LBM with 2nd-order $1/\rho$ approximation lies below $10^{-3}$, clearly showing that in this test case, the errors are dominated by the MPS decomposition.

Putting this in relation to the achieved compression, we find that at $\chi=64$ with a compression ratio of $\mathrm{CR}_{64}\approx120$, MPS-LBM maintains an accuracy of $95\%$ in terms of the inflow-outflow pressure difference. At a compression of $\mathrm{CR}_{128}\approx42$ with $\chi=128$ the deviation becomes negligible. 
Although the exact time evolution of the velocity field is not captured by compressed simulations, cases with $\chi \geq 64$ reproduce the pressure difference accurately, indicating that even with high compression rates, the underlying physics is preserved from startup to steady state.

\section{Discussion}\label{sec:discussion}

Quantum-inspired lattice Boltzmann based on matrix product state decomposition (MPS-LBM) addresses the memory bottleneck of LBM while it preserves high accuracy even for complex flow configurations. By compressing the MPS representation, MPS-LBM allows for substantial reductions in memory without altering the main algorithmic or grid structure. The method supports standard inflow/outflow boundary conditions and can handle arbitrary complex three-dimensional geometries within the computational domain, which was not demonstrated with MPS-based approaches before. Its capabilities are demonstrated through three 3D simulations designed to benchmark accuracy, flexibility, and compression performance.

The Taylor–Green vortex benchmark under varying compression ratios demonstrates that already at $\mathrm{CR}_{98}=64$, the temporal evolution of the kinetic energy is accurately reproduced. With $\mathrm{CR}_{256}=13$, fully resolved benchmark data are recovered. Complex geometries can be handled, as shown in the example of the flow in an aneurysm.
The compression ratio of the complex geometry mask imposes a lower bound on the achievable compression of the fluid domain, while only a modest increase in bond dimension beyond that of the mask is required to reproduce the flow accurately.
A pin-fin array geometry represents the practical application to a heat exchanger. Exploiting translational symmetry in the boundary mask, MPS-LBM achieves high data compression with accurate predictions. For instance at $\mathrm{CR}_{64}=120$, pressure drop error is less than $5\%$, while at $\mathrm{CR}_{128}=42$ reference data are recovered. \edit{While there is not a definitive way to determine a suitable choice of $\chi$ a priori, our results strongly indicate that the bond dimension necessary to accurately represent the geometry marks a lower bound. Apart from that further research is necessary to develop heuristics for the choice of $\chi$ or methods for automated adaptation.}
We anticipate that MPS-LBM delivers high computational performance in settings where the compressibility of translational symmetries (see \ref{sup:symmetry} for further details) can be exploited to reduce computational cost without compromising accuracy, thereby enabling simulations that would otherwise be infeasible due to computational constraints.

Due to the weakly compressible formulation, MPS-LBM does not require iterative solutions of a Poisson equation. Instead, it relies exclusively on element-wise operations within the MPS decomposition, which ensures that any algorithmic improvements to fundamental MPS operations translate directly into performance gains for MPS-LBM. 
The modular structure of the underlying LBM framework naturally supports algorithmic extensions, increasing the applicability of MPS-LBM. 
Heat transport~\cite{Karani_heat_2015}, including heat transfer, can be incorporated. Multi-relaxation-time schemes~\cite{dHumieres_mrt_2002} enable higher Reynolds numbers, while multiphase flow models~\cite{swift_multiphase_1996} extend the method to chemically relevant systems. Because the MPS decomposition introduces only minimal changes to the core LBM architecture, such extensions are straightforward while significantly expanding the range of MPS-LBM applications.

Another promising direction involves optimizing the decomposition of the binary geometry mask. Instead of minimizing the standard $\text{l}^2$-norm via singular value decomposition, alternative metrics such as the Hamming distance may be employed. Ideally, the decomposition scheme ensures that the contracted MPS mask remains binary. Advances in this area enable accurate simulations at even higher compression ratios, further extending the efficiency of MPS-LBM.

Finally, the lattice Boltzmann method has recently attracted attention as a candidate for fluid simulations on quantum computers~\cite{sanavio_carleman_2024, tiwari_algorithmic_2025, wawrzyniak_quantum_2025}. One of the primary challenges remains the treatment of collision, which is an active area of research~\cite{tennie_nonlinear_2025}. MPS-LBM, particularly the results presented in section B of the supplementary information, reduces this nonlinearity to element-wise multiplications with minimal loss of accuracy. More importantly, our results show that low-rank operations suffice to impose structured geometries onto LBM simulations, implying the feasibility of efficient implementations on quantum hardware.


\section*{Acknowledgments}
EM and LG would like to thank Hans Hohenfeld, Benedikt Placke, and Maximilian Kiefer-Emmanouilidis for helpful discussions. EM and LG acknowledge funding from the Federal Ministry for Economic Affairs and Energy (BMWE) and the German Aerospace Center (DLR) in the project QuMAL-KI under project number 50RA2208A, and from the Federal Ministry for Research Technology and Space (BMFTR) in the project QuaSA under project number 13N17300  administered by the VDI/VDE Innovation + Technik GmbH (VDI).
NAA acknowledges funding from ERC Advanced Grant Project No. 101094463. The funders played no role in study design, data collection, analysis and interpretation of data, or the writing of this manuscript. 

\section*{Author Contributions}
The project was conceptualized by DMW, JMW, EM, and NAA, and planned by LG, DMW, JMW, and EM. DMW and LG developed the approximation for inverse density. LG developed, implemented, and validated the software.
Methodological and implementation expertise for LBM was provided by DMW and JMW. Code review and minor development were conducted by DMW. Numerical experiments for the pin-fin heat sink case were planned by DMW and JMW, and those for the aneurysm and TGV cases by LG, DMW, JMW, and EM. LG set up and conducted all numerical experiments and performed data curation. LG and DMW performed post-processing of the data. The results were analyzed and interpreted by LG, DMW, JMW, and EM. DMW and JMW evaluated the physical consistency of the results. LG prepared the figures. The 3D flow visualizations were prepared by JMW and DMW. LG and DMW wrote the manuscript and the supplementary information, with major revisions by DMW. Revisions to the manuscript were conducted by JMW, EM, and NAA. The project was supervised by JMW, EM, and NAA.

\section*{Data Availability}
Due to size, the full experimental data will be made available upon reasonable request.

\section*{Code Availability}
The underlying code for this study is available at \cite{Gross_MPS-LBM_2026}

\section*{Competing Interest}
The authors declare no competing interests.

\section*{Declaration of generative AI and AI-assisted technologies in the manuscript preparation process.}
During the preparation of this work the authors used Microsoft Copilot and OpenAI ChatGPT to improve readability and language. After using these tools, the authors reviewed and edited the content as needed and take full responsibility for the content of the published article.

\appendix
\section{Two-Dimensional Test Cases}\label{sup:2dcases}
\begin{figure}[htb]
    \centering
    \includegraphics[width=.75\linewidth]{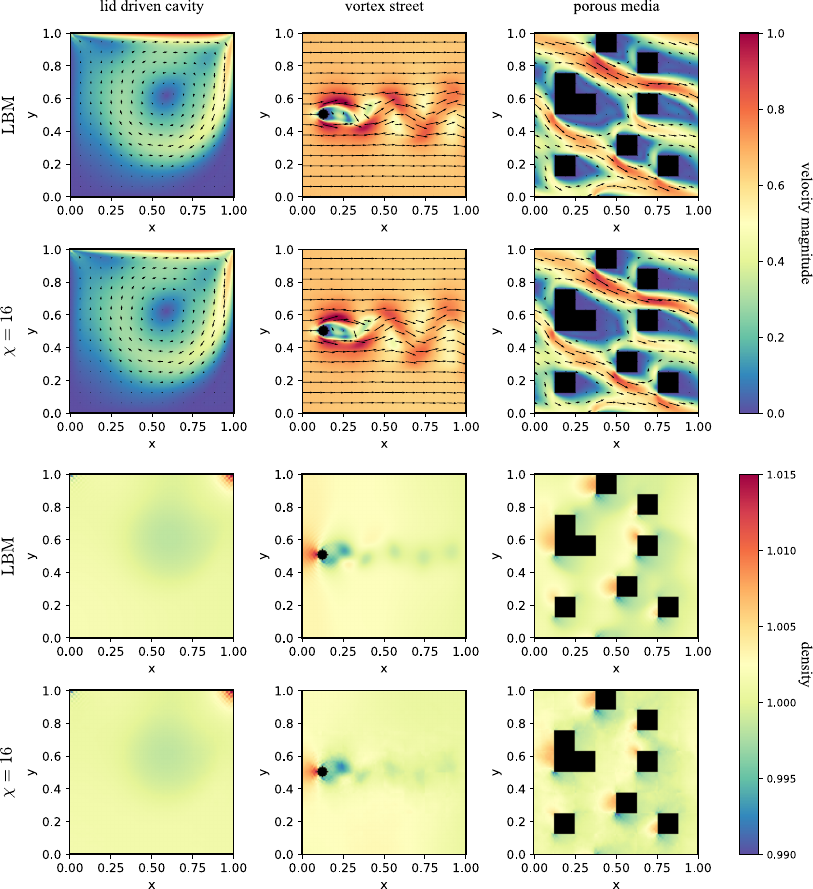}
    \caption{\textbf{Velocity magnitude and density of three common 2D test cases.} Shown from left to right: lid driven cavity, 
    vortex street, 
    and porous media.
    Top row shows standard LBM simulation, bottom row MPS-LBM at $\chi=16$ ($\text{CR} \approx 4.53$). Color map indicates velocity magnitude / density.}
    \label{fig:supp_experiments}
\end{figure}
We simulate three benchmark cases: lid-driven cavity, flow around a cylinder, and flow through a porous medium.
Each case is computed using both standard LBM and MPS-LBM algorithms at mesh resolutions of $128^2$.

All three scenarios are depicted in~\figref{fig:supp_experiments}. 
The bond dimension for all cases is $\chi=16$, which corresponds to a compression ratio of $\text{CR} = 4.53$.
The lid-driven cavity is evaluated at time $t=10\frac{L}{u_0}$ on a square domain with side length $L$. 
The fluid is initially at rest and driven by the top boundary moving at constant velocity, the Reynolds number is $\text{Re}=800$.
The flow around a cylinder of diameter $d$ is simulated at Reynolds number $\text{Re}=\frac{l_su_0}{\nu}=66$. 
The domain has edge length $L=16d$, and the velocity magnitude is shown at $t=\frac{5}{3}\frac{L}{u_0}$.
Finally, flow through a porous medium is driven by a constant volume force $(F_x,F_y)^T=(0.24,-0.1)^T~\si{\newton \per \meter^2}$.
Following the Shan–Chen forcing scheme \cite{shan_lattice_1993}, forces are incorporated via a perturbation term in the macroscopic velocity, which enters the equilibrium distribution function as
\begin{equation}
    \vec{u}=\frac{1}{\rho} \sum_if_i+\frac{\tau\vec{F}}{\rho},
\end{equation}
where we apply the same second-order approximation for $1/\rho$.
The maximum velocity reaches $u_{max}$, and the depicted velocity magnitude is shown at $t=5\frac{L}{u_\text{max}}$.
For all three scenarios from the reference, a bond dimension of $\chi = 16$ is sufficient to capture the flow dynamics with high fidelity. Only minor deviations are observed in the porous media and cylinder flow case.

\section{Approximation of Inverse Density}\label{sup:inverse}
\begin{figure}[htb]
    \centering
    \includegraphics[width=.6\linewidth]{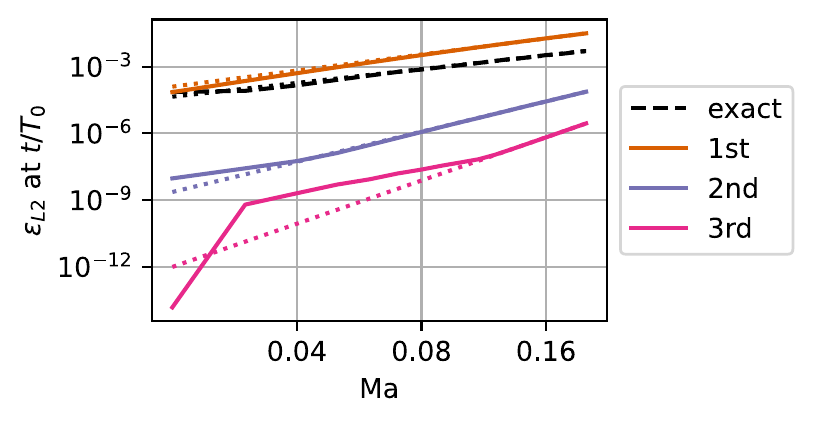}
    \caption{\textbf{Scaling of the $\text{l}^2$-error $\epsilon_{\text{l}^2}$ with Mach number Ma.} Results are from simulations of the 2D Taylor-Green vortex at different Ma. Solid colored lines show the error introduced by varying orders of approximation of $1/\rho$ with respect to the standard LBM. The dashed black line depicts the error of standard LBM with respect to the analytic solution of the TGV. Dotted lines indicate the respective linear fit.}
    \label{fig:inv_rho}
\end{figure}

We conducted a series of two-dimensional Taylor-Green vortex (TGV) simulations with edge length $L=2\pi$, characteristic velocity $u_0=1$, and initial conditions
\begin{equation}
    \begin{aligned}
        u(x,y) &= -  \sin\left(  x\right) \cos\left(  y\right), \\
        v(x,y) &=  \cos\left(  x\right) \sin\left(  y\right), \\
        \rho(x,y) &= 1 + \frac{1}{4\Delta u^2} \left(\cos\left(2 x \right) + \cos\left(2 y \right)\right).
    \end{aligned}
\end{equation}
The Reynolds number was set to $\text{Re} = \frac{u_0L}{\nu} = 125$. To maintain constant Re while varying Ma, the relaxation time was scaled as $\tau = 0.5 + \frac{N\text{Ma}}{c_s\ \text{Re}}$, with $N=256$ cells per spatial dimension.

The results obtained using Taylor expansions of $1/\rho$ up to third order were compared with standard LBM solutions and the analytical solution of the 2D TGV. 

In \figref{fig:inv_rho}, the colored lines show relative errors of the flow fields in the $\text{l}^2$-norm $\epsilon_{\text{l}^2}$ (see \cref{sec:metrics} in the main text) of LBM utilizing Taylor expansions of varying order for $1/\rho$ compared to the standard LBM in a log-log plot. The dashed black line indicates $\epsilon_{\text{l}^2}$ of the exact LBM compared to the analytic solution. Dotted lines show a linear fit to the respective data in log-log representation. 
Overall, the second-order approximation offers a favorable trade-off between computational efficiency and numerical accuracy relative to the standard approach.

\begin{table}
    \centering
    \caption{Fitted exponents of the scaling of $\epsilon_{\text{l}^2}$ with Mach number.}
    \label{tab:exp}
    \begin{tabular}{ccccc}
         & \textbf{Exact} & \textbf{1st} & \textbf{2nd} & \textbf{3rd} \\
        \hline
        \textbf{Exponent} & 2.059 & 2.401 & 4.513 & 6.457 \\
    \end{tabular}
\end{table}

The scaling exponents derived from these numerical experiments are summarized in \tabref{tab:exp}.
These were computed via a least squares method.
The function $\epsilon_{\text{l}^2}(\text{Ma},\alpha, \beta) = \alpha \text{Ma}^{\beta}$ was fitted to the computed set of data $(\epsilon_k, \text{Ma}_k)$ by minimizing the sum of squared residuals
\begin{equation}
    \min_{\alpha, \beta}\sum_i\left( \epsilon_k - \epsilon_{\text{l}^2}(\text{Ma}_k,\alpha, \beta)\right)^2.
\end{equation}

\section{Runtime Measurements}\label{sup:runtime}
\begin{figure}[htb]
    \centering
    \includegraphics[width=0.99\linewidth]{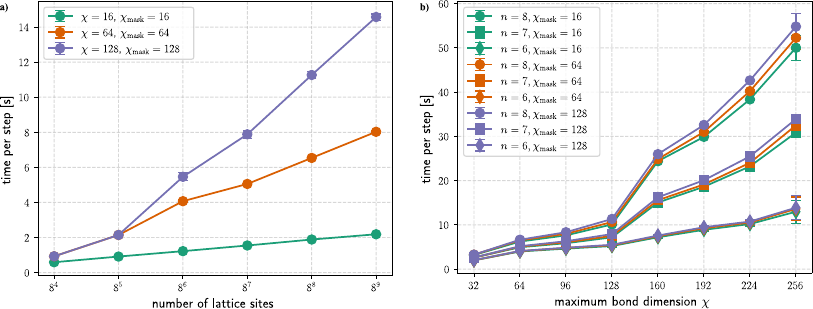}
    \caption{\textbf{Runtime scaling} of \textbf{a)} D3Q15 MPS-LBM with respect to the number of lattice sites  \textbf{b)} and the maximum bond dimension.}
    \label{fig:runtime}
\end{figure}

In this subsection, we provide data on measured runtimes and discuss them. For all measurements, we performed $100$ time steps of the D3Q15 MPS-LBM to compute the mean and standard deviation. In- and outflow conditions, as well as a geometry mask, are enabled. Tensor manipulations were implemented using the Python packages JAX~\cite{jax2018github} and opt\_einsum~\cite{Smith_opteinsum_2018} and all simulations were conducted on an NVIDIA A100 GPU with 80GB of VRAM.

In \subfigref{fig:runtime}{a}, the scaling with respect to the number of lattice sites $N=8^n$ is shown, where $n$ is the number of sites in each MPS for different bond dimension $\chi \in \{ 16,64,128\}$. The expected linear scaling in $n$ is clearly visible. The only deviation at $N=8^4$ and $N=8^5$ for $\chi = 128$ is explained by the fact that MPS of these sizes are already exact at a bond dimension of $64$, and MPS-LBM hence operates on MPS smaller than the allowed maximum bond dimension of $\chi = 128$.

The scaling with the bond dimension, shown in \subfigref{fig:runtime}{b}, is more complex. The scaling with $\chi^4$ cannot clearly be discerned due to an inconsistency between $\chi=128$ and $\chi=160$, as well as between $\chi=224$ and $\chi=256$. Presumably, these irregularities originate from GPU saturation effects or optimization decisions made by the compiler of JAX~\cite{jax2018github}.

\begin{figure}[htb]
    \centering
    \includegraphics[width=0.99\linewidth]{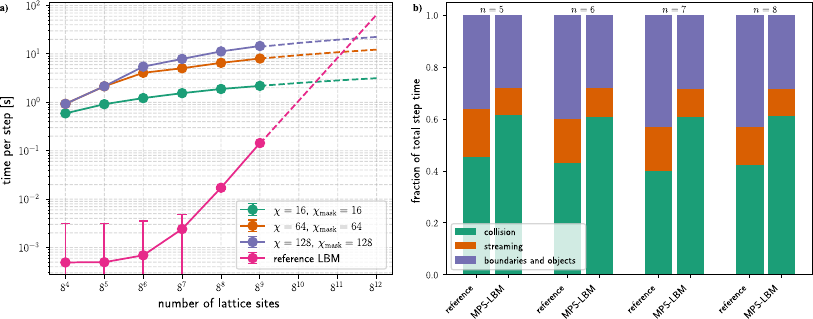}
    \caption{\edit{\textbf{Runtime comparison with the reference LBM.} \textbf{a)} Time in seconds per step including extrapolations (dashed lines) of the runtime crossover of MPS-LBM and the reference. \textbf{b)} Comparison of the fraction of time spent in the collision, streaming and the handling of obstacles and boundaries.}}
    \label{fig:runtime_comp}
\end{figure}
\edit{
As comparison with the reference LBM \subfigref{fig:runtime_comp}{a} shows runtimes on a log-scale. Initially the reference LBM is several orders of magnitude faster. After saturation effects of the GPU below $N=8^7$, the data exhibits the expected exponential growth. Dashed lines show the extrapolation of our data beyond $N=8^9$. Given our implementation the data suggest that for moderate $\chi$, MPS-LBM exhibits a runtime improvement at system sizes larger than $8^{11}$. It is important to note that this observation depends on specific implementation and hardware. 

Additionally, in \subfigref{fig:runtime_comp}{b} the fraction of time spent on the collision step, the streaming step and the handling of boundaries and obstacles respectively, is shown for various $n$. In the reference LBM collision and handling of obstacles and boundaries have a similar cost, while the streaming takes roughly half as much time. In MPS-LBM the cost of the collision increases compared to the reference, marking this operation a candidate for future improvement. These observations do not vary significantly with varying $n$.
}

\section{Symmetric Geometries in Matrix Product State Decomposition}\label{sup:symmetry}
\begin{figure}[htb]
    \centering
    \includegraphics[width=0.99\linewidth]{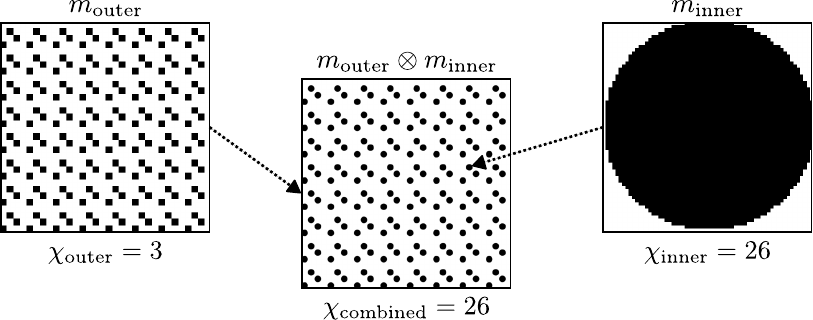}
    \caption{\textbf{Combination of masks on separate length scales.}}
    \label{fig:symmetry}
\end{figure}
We examine the relation between MPS decomposition and translationally symmetric geometries using an explicit example, and show that the approach generalizes to realistic cases. Consider a geometry with separation of large- and small-scale structures.
We consider an inner geometry represented as a binary mask $m_{\mathrm{inner}}$ of resolution $2^{kD}$, repeated according to an outer mask $m_{\mathrm{outer}}$ on a $2^{lD}$ grid. The combined mask inserts the inner geometry at each site where $m_{\mathrm{outer}} = 1$, leaving other sites zero.
An example is shown in \figref{fig:symmetry}, where the inner mask is a circle on a $64^2$ grid, arranged in the repetition pattern defined by an outer mask on a $32^2$ grid. For simplicity, in 1D the value at position $x$ is given by
\begin{equation}
    m_{\mathrm{combined}}^x = m_{\mathrm{outer}}^{\left\lfloor x / 2^k \right\rfloor} \ m_{\mathrm{inner}}^{x \bmod 2^k } = (m_{\mathrm{outer}} \otimes m_{\mathrm{inner}})^x.
\end{equation}
Thus, it is straightforward to see that the combined mask is in fact the tensor product of the outer and inner masks.
To relate this structure to MPS, we examine the index representation. Consider the scale-ordered decomposition
\begin{equation}
    \begin{aligned}
                  &\vec{x} = x_1 \dots x_D = b_1^1 \dots b_1^n \dots b_D^1 \dots b_D^n \\
        \to \quad &b_1^1 \dots b_D^1 \dots b_1^n \dots b_D^n = w_1 \dots w_n,        
    \end{aligned}
\end{equation}
as used in \cref{sec:mps} of the main text. Let us denote the scale ordered indices of $m_{\mathrm{inner}}$ and $m_{\mathrm{outer}}$ as $\vec{x}_{\mathrm{in}} \to u_1 \dots u_k$ and $\vec{x}_{\mathrm{out}} \to v_1 \dots v_l$ and consider both masks to be MPS decomposed, then we can write out the tensor product as
\begin{equation}
    \begin{aligned}
        (m_{\mathrm{outer}} \otimes m_{\mathrm{inner}})^{\vec{x}_{\mathrm{out}} \vec{x}_{\mathrm{in}}}
        &= m_{\mathrm{outer}}^{v_1 \dots v_l}  m_{\mathrm{inner}}^{u_1 \dots u_k} \\
        &= \sum_{\{\vec{\alpha}\},\{\vec{\beta}\}} \left(\mat{A}_1\right)_{\alpha_1}^{v_1} \dots \left(\mat{A}_l\right)_{\alpha_{l-1}}^{v_l} \ \left(\mat{B}_1\right)_{\beta_1}^{u_1} \dots \left(\mat{B}_k\right)_{\beta_{k-1}}^{u_k}.
    \end{aligned}
\end{equation}
Therefore, the maximum bond dimension of the combined mask equals the larger of the two masks $\chi_{\mathrm{combined}} = \max (\chi_{\mathrm{outer}}, \chi_{\mathrm{inner}})$.
In the example shown in \figref{fig:symmetry}, the combined mask has a resolution of $2048^2$ and is accurately represented as an MPS with $\chi_{\mathrm{combined}} = 26$.
Together with the results from our simulations, this suggests that MPS-LBM would produce physically accurate flows around $\chi\approx 64$ ($\mathrm{CR}\approx46$).
This construction directly relates to the pin-fin scenario, where the inner mask defines a single pin-fin and the outer mask specifies the array layout. The concept generalizes to other applications.

\bibliographystyle{elsarticle-num} 
\bibliography{references}

\end{document}